\documentclass[journal]{vgtc-no-name}

\renewcommand{\manuscriptnotetxt}{}
\nocopyrightspace

\newcommand{\tool}{\textsl{TopoAlign}}

\usepackage{amssymb}
\usepackage{amsmath}
\usepackage{enumitem} 
\usepackage{booktabs}

\newcommand{\mm}[1]{\ifmmode{#1}\else{\mbox{\(#1\)}}\fi}

\newcommand{\Rspace}{\mm{\mathbb{R}}}
\newcommand{\Xspace}{\mm{\mathbb{X}}}
\newcommand{\Ucal}{\mm{\mathcal{U}}}
\newcommand{\Vcal}{\mm{\mathcal{V}}}

\newcommand{\para}[1]{\noindent\textbf{#1}}

\vgtccategory{Research}

\title{\textbf{TopoAlign}: Topology-Aware Visual Representation Alignment}

\author{
  \authororcid{Xinyuan Yan}{0000-0003-3396-1310}, 
  \authororcid{Rita Sevastjanova}{0000-0002-2629-9579}, 
  \authororcid{Mennatallah El-Assady}{0000-0001-8526-2613},
  and \authororcid{Bei Wang}{0000-0002-9240-0700}
}

\authorfooter{
\item Xinyuan Yan and Bei Wang are with University of Utah.\\
E-mails: \{xinyuan.yan, beiwang\}@sci.utah.edu
\item Rita Sevastjanova and Mennatallah El-Assady are with
ETH Zürich.\\
E-mails: rita.sevastjanova@inf.ethz.ch, 
menna.elassady@ai.ethz.ch.
}

\abstract{
Neural networks encode inputs as high-dimensional vectors, known as \emph{representations}, that capture how models process data by encoding task-relevant structure and semantics. Representation alignment refers to the degree to which different models, layers, or training conditions produce similar representations for the same inputs, with important implications for model interpretation, selection, and robustness analysis. Existing approaches to measure alignment primarily rely on geometric properties, such as neighborhood and cluster similarity, offering limited insight into the global organization of representations.
In this work, we present {\tool}, a topology-aware framework for visually comparing model representations from a structural perspective. Leveraging mapper graphs from topological data analysis, {\tool} jointly analyzes graphs constructed from representations of shared inputs across different models or layers. The framework supports a top-down comparative workflow: it first performs global structure alignment via joint force-directed optimization to produce coordinated graph layouts; it then identifies local correspondences through automated detection of structurally matching regions, visualized with Bubble Sets; and finally it enables fine-grained pattern inspection through motif-based queries and membrane-inspired visualizations.
We demonstrate {\tool} through case studies on language and multimodal models, complemented by expert feedback. Our results show that {\tool} provides meaningful insights into representation structure and alignment from a topological perspective.
}


\teaser{
\centering
\includegraphics[width=\linewidth]{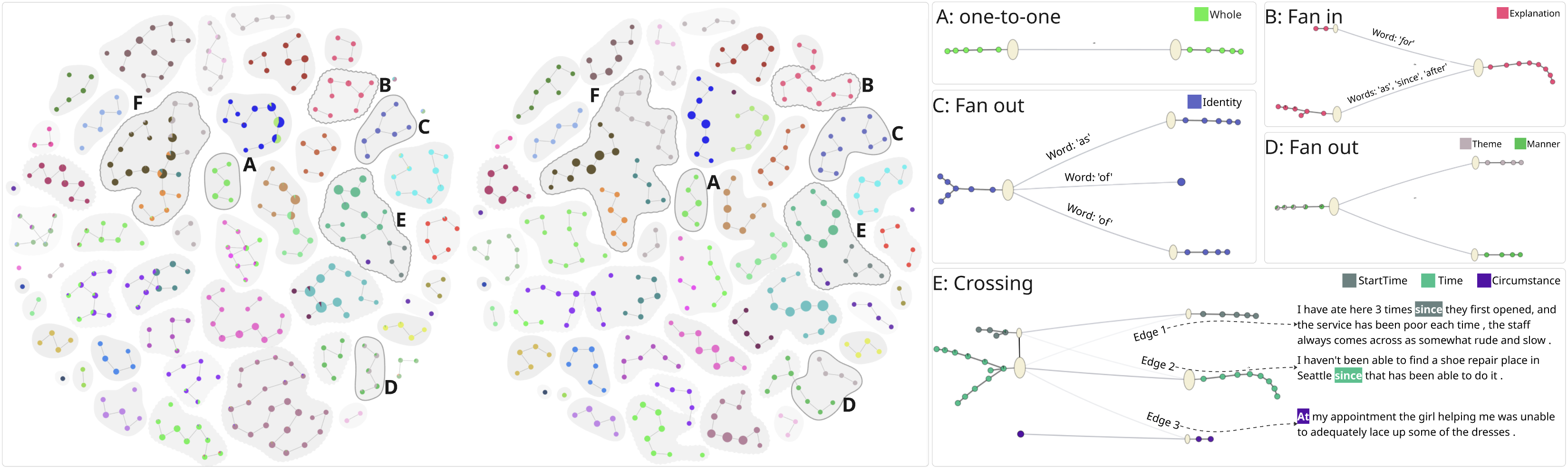}
\caption{Topology-aware visual representation alignment during fine-tuning with {\tool}.
Left: Globally aligned mapper graphs of word representations extracted from the final layer of fine-tuned BERT-Base at epochs 2 and 6. Local alignments—computed with priority assigned to the topology of the epoch-2 mapper graph—are visualized using Bubble Sets. Alignment groups (A–F) are highlighted by boundary contours.
Right: Membrane views of selected alignments (A–E), showing the corresponding local structures in the two mapper graphs (epoch 2 vs.\ epoch 6). Inter-edges indicate shared items and reveal their alignment relationships.}
\label{fig:BERT-epoch-compare}
}

\begin{document}

\maketitle

\section{Introduction}
\label{sec:introduction}

Neural network models have become increasingly pervasive, transforming a wide range of data modalities, such as text and images, and powering diverse applications, ranging from conversational agents~\cite{singh2025survey} to autonomous systems~\cite{atakishiyev2024explainable} to scientific discovery~\cite{reddy2025towards}.
To understand and improve model behavior, model representations---high-dimensional vectors encoding how inputs are represented within a model---have attracted significant attention due to their ability to shed light on the internal workings of neural networks~\cite{boggust2022embedding,rissom2024decoding,yan2025explainable, sevastjanova2021explaining,sivaraman2022emblaze}.

A fundamental task in representation analysis is representation alignment, which measures how similarly two models, or different layers of the same model, represent a shared set of inputs~\cite{sucholutsky2025getting}.
Such analysis have broad implications, including understanding model learning dynamics across layers and training stages~\cite{mehrer2018beware,morcos2018insights}, informing model architecture design~\cite{nguyen2021do,gwilliam2022beyond}, guiding model selection for downstream tasks~\cite{rissom2024decoding,sivaraman2022emblaze}, and improving robustness~\cite{jones2022if,nanda2022measuring}.

To quantify alignment, the machine learning community has proposed numerous metrics~\cite{klabunde2025resi, klabunde2025similarity}, such as pairwise similarity measures~\cite{kriegeskorte2008representational} and neighborhood-based comparisons~\cite{gwilliam2022beyond}.
However, these aggregated measures are often overly simplistic and provide limited interpretability~\cite{klabunde2025similarity}.
To address this limitation, visual analytics approaches have been developed to reveal fine-grained alignment patterns.
Some approaches~\cite{boggust2022embedding,sivaraman2022emblaze,heimerl2020embcomp} decompose global metrics to highlight individual data points whose neighborhood change the most or least across models, while others~\cite{arendt2020parallel,yu2026parallel} cluster representations and establish correspondences between clusters for scalable comparisons. 
Despite these advances, existing approaches are primarily grounded in geometric notions of similarity that rely on pointwise distances.

In contrast, topology~\cite{carlsson2009topology} provides a complementary perspective by capturing the global shape and organization of data. For example, the mapper graph~\cite{SinghMemoliCarlsson2007}, a widely used technique in topological data analysis (TDA), abstracts a high-dimensional point cloud into a graph, where nodes represent clusters and edges denote overlaps between them. Mapper has been successfully applied to analyze model representations across domains such as language models~\cite{rathore2023topobert,yan2025explainable,rair2025annotators}, computer vision~\cite{rathore2021topoact,purvine2023experimental}, and adversarial attacks~\cite{zhou2023visualizing}, revealing meaningful structural patterns in representation spaces through elements such as nodes, paths, and components~\cite{yan2025explainable}. However, prior work primarily uses mapper to analyze a single model in isolation or to manually compare multiple mapper graphs, and therefore lacks automated and systematic methods for comparing topological structures across models.

To bridge this gap, we propose a systematic approach for comparing the topological structures of model representations via mapper graphs. Given two mapper graphs constructed from the same input data under different models, layers, or training conditions, our method performs comparison in a multi-scale, top-down manner. 
At the \textbf{global level}, we analyze coordinated graph layouts to expose differences in overall structural organization. At the \textbf{local level}, we identify and compare structurally corresponding regions (i.e., subgraphs) that capture how subsets of data are organized across representations. At a \textbf{finer scale}, we further examine detailed alignment patterns within these regions to characterize how features (e.g., clusters) are preserved, split, or merged.

To support this analysis, we introduce {\tool}, a topology-aware visual analytics framework that enables machine learning experts to analyze representation alignment across models. By jointly modeling and visualizing pairs of mapper graphs, {\tool} makes the following novel contributions:
\begin{itemize}[noitemsep,leftmargin=*]
\item \textbf{Globally aligned mapper visualizations:} A joint graph layout optimization framework that produces coordinated embeddings, enabling direct comparison of global topological structures.
\item \textbf{Automatic cross-graph region correspondence:} An alignment-aware clustering approach that identifies structurally corresponding subgraphs between mapper graphs.
\item \textbf{Interpretable local alignment characterization:} A Bubble Sets-based visualization coupled with glyph-driven querying to reveal local alignment patterns and assess alignment quality.
\item \textbf{Multi-scale alignment inspection:} A membrane-inspired visualization technique that supports detailed exploration of fine-grained structural correspondences.
\item \textbf{Integrated interactive system:} A unified visual analytics workspace that integrates these components to support exploratory analysis of representation alignment.
\end{itemize}
We demonstrate the effectiveness of {\tool} through multiple use cases on language and multimodal models. Our approach reveals meaningful structural phenomena---including feature splitting, merging, and disentanglement---across training stages, layers, models, and modalities. Preliminary expert feedback suggests that {\tool} supports insight generation and highlights opportunities for further investigation. To the best of our knowledge, this is the \emph{first} work to systematically compare neural network representations using topological abstractions.

\section{Related Work}
\label{sec:related-work}

\subsection{Representation Alignment in Machine Learning}
Representation alignment~\cite{sucholutsky2025getting,klabunde2025similarity} is a fundamental task in machine learning (ML) community, studying how similarly different models, layers, or training stages encode the same input data. 
It plays a key role in understanding model behavior~\cite{nguyen2021do,gwilliam2022beyond,edamadaka2025universally}, analyzing learning dynamics~\cite{mehrer2018beware,morcos2018insights}, and assessing robustness and generalization~\cite{jones2022if,nanda2022measuring}.
A large body of work measures alignment through representation similarity metrics~\cite{klabunde2025similarity,klabunde2025resi}. 
Given two models and a shared input dataset, these methods compute a scalar similarity score between their model representations. 
One common approach~\cite{kriegeskorte2008representational,shahbazi2021using,kornblith2019similarity} computes pairwise similarity matrices within each space and measuring their differences. 
Some other methods ~\cite{ding2021grounding,williams2021generalized} align representations directly by learning transformations that minimize discrepancies between the two spaces. 
Additionally, neighborhood-based approaches~\cite{schumacher2021effects,wang2020towards, hryniowski2020inter} compare the Jaccard similarity of $k$-nearest neighbors for each data point. 
Comprehensive surveys~\cite{klabunde2025similarity}, benchmark studies~\cite{romero2024resilient}, and systematic evaluations~\cite{wu2025measuring} have assessed these metrics, consistently showing that no single one captures all aspects of representational similarity. 

However, these methods primarily produce aggregated measures, often obscuring nuanced structural patterns. In contrast, we model representations as mapper graphs and visually analyze their alignment to enable multi-scale comparisons that reveal both global and local structures.

\subsection{Visualization for Embedding Comparison}

Several tools support language model comparison by enabling inspection of contextual word embeddings.
EmbeddingVis~\cite{li2018embeddingvis} is an exploratory visual analytics system for comparing embedding vectors at cluster, instance, and structural levels, and for assessing node metric preservation across models.
embComb~\cite{heimerl2020embcomp} uses various metrics to quantify differences in the local structure around embedding objects.
Arendt et al.~\cite{arendt2020parallel} support concept-oriented model comparison, revealing qualitative differences in how models interpret input data.
Embedding Comparator~\cite{boggust2022embedding} facilitates embedding comparison via small multiples. 
It calculates and displays similarity scores for embedded objects based on their local neighborhoods (i.e., shared nearest neighbors). 
Emblaze~\cite{sivaraman2022emblaze} uses an animated scatterplot and integrates visual augmentations to summarize changes within the embedding spaces under study. 
LMFingerprints~\cite{sevastjanova2022lmfingerprints} employs scoring methods to analyze the properties captured in embedding vectors, allowing comparisons between different models as well as between layers of the same model.
In another work, Sevastjanova et al.~\cite{sevastjanova2022adapters} also compare adapter models to investigate potential biases present in word embedding spaces.

Comparison-focused visualizations are used across multiple data modalities, including image-to-text setup. 
Notable examples include VAC-CNN~\cite{xuan2022vac}, which enables detailed inspection and comparison of CNN models, and DKMap~\cite{ye2026dkmap}, which supports image–text alignment analysis in vision-language models.
Unlearning Comparator~\cite{lee2026unlearning} system is designed to compare different machine unlearning models (i.e., models that can selectively forget specific classes). 
The authors use two image datasets for image classification, derive image embeddings, and apply UMAP for projection, providing both class-level and layer-wise views.
The most recent work in this domain is ParaClus~\cite{yu2026parallel}. 
It supports multi-scale exploration of embedding structures and attributes using adaptive thresholding to delineate local neighborhoods.

However, these works focus on the geometry of representations, whereas {\tool} offer a topological view for comparative analysis.

\subsection{Mapper Graph for High dimensional Data Analysis}

Mapper graphs~\cite{SinghMemoliCarlsson2007} have found applications across a wide range of domains, including cancer 
research~\cite{NicolauLevineCarlsson2011,MathewsNadeemLevine2019}, sports analytics~\cite{Alagappan2012}, gene expression analysis~\cite{JeitzinerCarriereRougemont2019}, 
micro-epidemiology~\cite{Knudson2020}, and neuroscience~\cite{GeniesseSpornsPetri2019,SaggarSpornsGonzalez-Castillo2018}, with a comprehensive survey available in~\cite{PataniaVaccarinoPetri2017}. 
Recently, mapper graphs have become increasingly important in explainable AI. 
They have been used to analyze representations from image classifiers and large language models~\cite{rathore2021topoact,rathore2023topobert,rair2025annotators,yan2025explainable}, to study how representation spaces evolve across layers~\cite{purvine2023experimental} and during fine-tuning~\cite{rathore2023topobert}, and to examine patterns induced by adversarial attacks~\cite{ZhouZhouDing2023}.
In addition,
GALE~\cite{xenopoulos2022gale} models the relationship between explanation spaces and model predictions as a scalar function, using its mapper graph as a structural signature to consistently compare explanation methods.
MOUNTAINEER~\cite{solunke2024mountaineer} builds on this idea by providing an interactive visual analytics framework for comparing heterogeneous local explanations. However, these work either focuses on single mapper graphs or relies on manual inspection of node correspondences for comparison.

{\tool} similarly leverage mapper graphs as a unified topological abstraction of high-dimensional representation spaces, but goes beyond by introducing a systematic framework for both global and local comparison, aided by visual encodings to interpret alignment patterns.

\section{Technical Background}
\label{sec:background}

\subsection{Mapper Graphs}
\label{sec:mapper-graphs}

Mapper graphs provide a principled framework for summarizing high-dimensional representation spaces by organizing points into overlapping local clusters and encoding their connectivity as a graph~\cite{SinghMemoliCarlsson2007}. This construction preserves both local neighborhood structure and global topology, yielding a coherent, multi-scale view of learned representations that is well-suited for analysis and comparison.

\begin{figure}[!ht]
\centering
\includegraphics[width=0.8\linewidth]{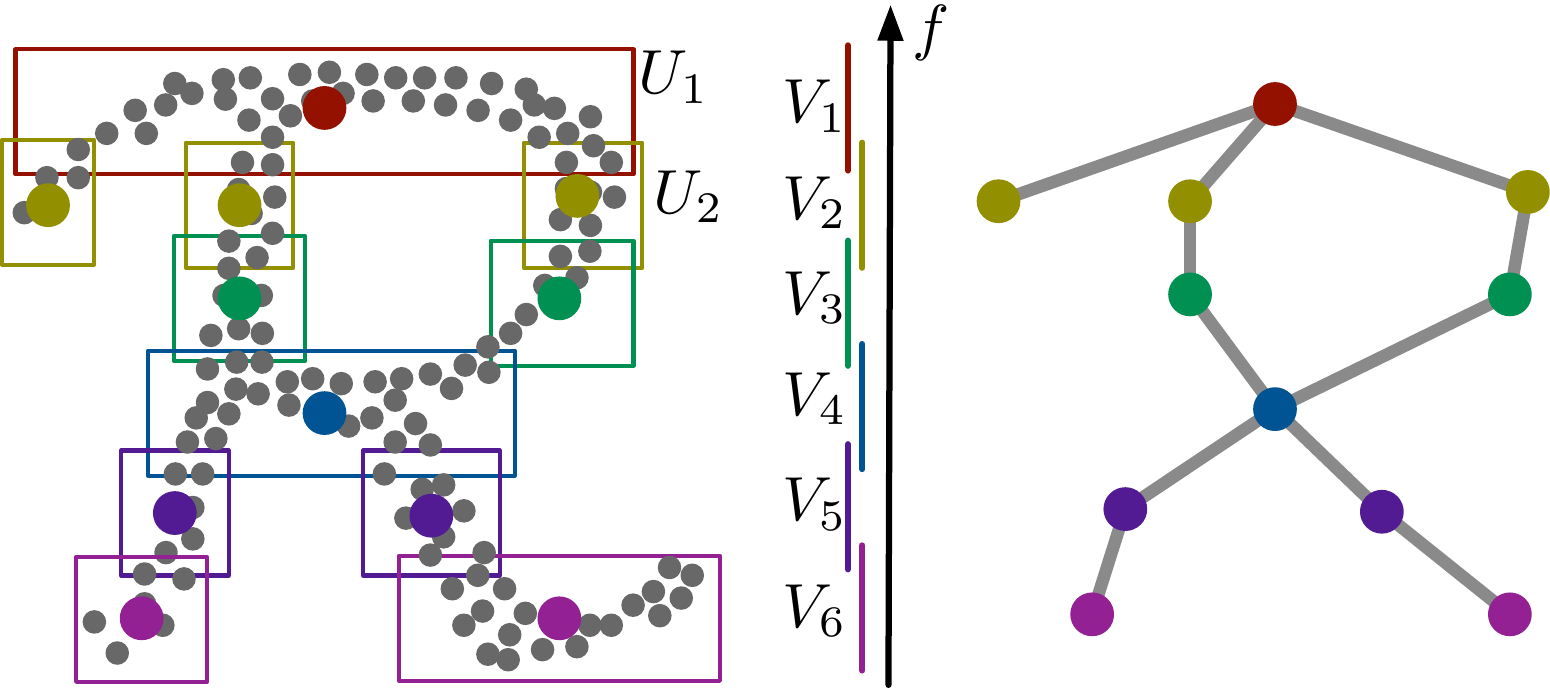}
\vspace{-2mm}
\caption{An illustration of mapper graph construction. Left: a point cloud $\Xspace$ equipped with a height function $f$. Middle: a cover $\Ucal = \{U_1, U_2, \dots\}$ of $\Xspace$ induced by an overlapping cover $\Vcal = \{V_1, V_2, \dots\}$ of $f(\Xspace)$, consisting of six intervals with $25\%$ overlap. Right: the one-dimensional nerve of $\Ucal$ produces the mapper graph.}
\label{fig:mapper}
\vspace{-4mm}
\end{figure}

Formally, given a point cloud $\Xspace$ of latent representations and a \emph{filter function} $f: \Xspace \to \Rspace$, the mapper algorithm constructs a graph whose nodes correspond to clusters and whose edges indicate nonempty intersections between them. To illustrate, consider a point cloud $\Xspace$ sampled from a shape resembling the letter $R$ (see~\cref{fig:mapper}). Equipping $\Xspace$ with a filter function $f$ (e.g., a norm capturing variation in representations), we define an overlapping cover $\Vcal = \{V_1, V_2, \dots\}$ of $f(\Xspace)$ such that $f(\Xspace) \subseteq \bigcup_j V_j$. Each interval $V_j$ induces a subset $f^{-1}(V_j)$, which is clustered to produce a cover $\Ucal = \{U_1, U_2, \dots\}$ of $\Xspace$; for example, $f^{-1}(V_2)$ may yield three clusters, while $f^{-1}(V_3)$ yields two.

The mapper graph is defined as the one-dimensional \emph{nerve} of $\Ucal$: each cluster $U_i$ becomes a node, and edges connect pairs $(i,j)$ whenever $U_i \cap U_j \neq \emptyset$. In the spirit of soft clustering, allowing overlap between clusters is crucial for preserving continuity across neighboring regions, which is essential for robust alignment analysis.

In our setting, we use the $\ell^2$-norm as the default filter function, which measures the magnitude of representation vectors and has been shown to reveal meaningful structure in deep models~\cite{rathore2021topoact,rathore2023topobert}. With $f$ fixed, the mapper construction is governed by: (1) the number and overlap of intervals in $\Vcal$, controlling resolution; and (2) the clustering procedure. We employ DBSCAN~\cite{ester1996density}, with parameters $\epsilon$ (neighborhood radius) and $\texttt{minPts}$ (minimum cluster size). 

By encoding how local regions overlap, the mapper graph captures the global organization of the representation space (e.g., branching or loops in the letter $R$). Nodes represent semantically coherent regions, and edges represent transitions between them. This makes the representation well-suited for alignment: similarities between spaces can be assessed via correspondences in graph topology, cluster structure, and connectivity across models, layers, or modalities.

\subsection{Force-Directed Graph Layouts}
\label{sec:background_fd}
Force-directed layout~\cite{fruchterman1991graph, kamada1989algorithm, hu2005efficient} is a widely used technique for visualizing graphs by simulating a physical system, where nodes are treated as particles and edges as springs. 
Given a graph \(G = (V, E)\), where \(V\) is the set of nodes and \(E \subseteq V \times V\) is the set of edges, force-directed layout seeks to compute 2D positions
$\mathbf{p} = \{\, p_v \in \mathbb{R}^2 \mid v \in V\,\}$
for all nodes $v \in V$ by minimizing an energy function that balances attractive and repulsive forces:
\[
\mathcal{L}(\mathbf{p}) = \sum_{(u,v) \in E} \left( \|p_u - p_v\| - \ell \right)^2 
+ \sum_{\substack{u,v \in V \\ u \neq v}} \frac{k}{\|p_u - p_v\|}.
\]
The first term encourages adjacent nodes \((u,v)\in E\) to be placed at a preferred edge length \(\ell\), while the second term imposes repulsion between all pairs of nodes, with constant \(k>0\), to reduce overlap and improve readability. 
This energy is typically minimized using iterative methods, where node positions are updated according to the induced forces until convergence. 
The resulting layouts effectively reveal both local connectivity and global structure, leading to their widespread adoption across various domains~\cite{cheong2020force}.
Besides, various extensions have been proposed to improve its scalability~\cite{zhong2023force}, reduce visual clutter~\cite{doppalapudi2022untangling}, or adapt the layout to meet user-specified criteria~\cite{xue2025autofdp}. 
We extend this formulation by introducing additional forces to align two mapper graphs both globally and locally (see \cref{sec:alignment_layout,sec:multiscale_alignment}).

\section{Task Analysis and Design Requirements}
\label{sec:task_and_requirements} 

This work aims to enable ML experts---familiar with model architectures and representations---to visually analyze representational alignment from a topological perspective. Given two mapper graphs constructed from shared inputs under different models (or layers), we investigate: (1) what aspects should be compared and how they inform topological and representation analysis (\cref{sec:task}); and (2) the resulting visualization design requirements (\cref{sec:requirements}). Our analysis is informed by long-term collaboration with ML researchers, the expertise of a TDA specialist on our team, and an extensive review of the literature.

\subsection{Comparative Task Analysis}
\label{sec:task}
Correspondence between two mapper graphs can be established via node membership, by tracking how data points from a node in one graph are distributed across nodes in the other. Using this, we perform a top-down comparison from global to local analysis.

\para{T1. Global structure comparison.}
At the global level, we compare the overall shape of the two mapper graphs to obtain a high-level overview of similarities and differences between the model representation spaces.
From a TDA perspective, this global shape reflects the coarse topology of the data through connectivity patterns~\cite{rathore2023topobert,rathore2021topoact,solunke2024mountaineer,rair2025annotators}.
Comparing these structures reveals the extent to which the models exhibit similar global organization of their representation spaces.

\para{T2: Local correspondence comparison.}
At the local level, structural elements in mapper graphs, such as nodes, paths, and components, encode meaningful patterns of data organization~\cite{yan2025explainable}. 
For example, components may represent a semantically coherent cluster, while a path can indicate continuous transitions in the representation space. 
Establishing correspondences between such local structures across graphs helps identify where models agree or diverge in organizing specific subsets of data. 

\para{T3: Fine-grained pattern inspection.}
Given a pair of local correspondences, alignment patterns can take different forms.
For instance, one-to-one correspondences between components indicate that data features are represented consistently across models, whereas one-to-many correspondences suggest that a feature is refined or decomposed in another model.
From a TDA perspective, these patterns reflect concepts such as feature (e.g., cluster) persistence, splitting, and merging~\cite{carlsson2009topology}.
Analyzing these patterns offers deeper insights into how models differ in organizing and representing knowledge.

\subsection{Visualization Design Requirements}
\label{sec:requirements}
We derive the following design requirements to support tasks \textbf{T1}-\textbf{T3}.

\para{R1. Globally aligned visualization of mapper graphs.}
To support global comparison (\textbf{T1}), the system should present the two mapper graphs in a visually aligned layout. 
While side-by-side arrangements facilitate comparison~\cite{gleicher2011visual}, mapper graphs are often independently generated via force-directed layouts, resulting in inconsistent spatial correspondence.
A key challenge is therefore to align the layouts so that similar regions regarding the underlying data are positioned comparably across graphs, enabling effective visual comparison.

\para{R2. Interactive and automated discovery of local alignments.}
To support local comparison (\textbf{T2}), the system should enable both user-driven and automated discovery of local alignments between mapper graphs. 
\para{R2-1.}~Users can explore alignments by interactively selecting elements of interest (e.g., a subgraph) in one graph and examining their counterparts in the other.
However, such manual exploration can be time-consuming in identifying meaningful alignments and may overlook important patterns. 
\para{R2-2.} The system should also provide automated methods to identify local alignments and characterize their properties, offering a more efficient analysis basis.

\para{R3. Characterization of local alignment patterns.}
To support interpretation of local alignment (\textbf{T3}), the visualization should provide high-level summaries of alignment patterns.
This characterization enables users to quickly understand the implications of local alignments, supporting efficient reasoning about model differences.

\para{R4. Support for multi-scale, fine-grained inspection.}
To enable detailed inspection of local alignments (\textbf{T3}), the system should support multi-scale exploration of node correspondences.
Directly drawing edges between corresponding nodes across the two subgraphs can create severe visual clutter due to edge crossings and density.
Therefore, the visualization should allow exploration of local alignments at multiple levels of abstraction while minimizing  visual clutter.
\section{Method: Visual Representation Alignment}
\label{sec:method}

In this section, we present \textit{TopoAlign}, a topology-aware approach for visually analyzing representation alignment between neural network models, designed to satisfy the requirements \textbf{R1}--\textbf{R4}.

\para{Input and problem setup.}
Let $\mathcal{D} = \{x_i\}_{i=1}^n$ denote a set of shared input data. 
Given two neural network models $f_1$ and $f_2$, we extract their representations at selected layers:
\[
\mathbf{Z}_1 = f_1(\mathcal{D}) \in \mathbb{R}^{n \times d_1}, \quad
\mathbf{Z}_2 = f_2(\mathcal{D}) \in \mathbb{R}^{n \times d_2},
\]
where $n$ is the number of data points, and $d_1$, $d_2$ are the representation dimensions. 
We then construct two mapper graphs from the representations using identical parameter configurations to ensure comparable resolution: $G_1 = (V_1, E_1)$ and $G_2 = (V_2, E_2)$.
Each node $v \in V_k$ represents a subset of data points, while an edge $(u,v) \in E_k$ indicates overlapping membership between the corresponding nodes. These two graphs serve as the basis for our comparative analysis. 
Although we describe the setting in terms of comparing two models, the formulation naturally extends to comparisons across different layers, training stages, or modalities within a single model.

\para{Approach overview.}
As the foundation of our approach, we construct a joint mapper graph $G_{\text{joint}}$ by introducing edges between nodes in $G_1$ and $G_2$ that share common data points, thereby explicitly encoding correspondences across the two graphs:
\[
G_{\text{joint}} = (V_1 \cup V_2,\; E_1 \cup E_2 \cup E_{\text{inter}}).
\]

The inter-graph edge set $E_{\text{inter}}$ is defined based on shared data memberships. Specifically, for nodes $u \in V_1$ and $v \in V_2$, an inter-graph edge is introduced when their associated data subsets overlap:
\[
E_{\text{inter}} = \{(u, v) \mid u \in V_1,\; v \in V_2,\; S_u \cap S_v \neq \emptyset \},
\]
where $S_u$ and $S_v$ denote the sets of data points associated with nodes $u$ and $v$, respectively.

Built upon the joint mapper graph, we propose a force-directed optimization framework that produces globally aligned layouts, preserving the internal structure of each graph while spatially aligning corresponding regions (\textbf{R1}; \cref{sec:alignment_layout}). 
We then cluster the joint graph to automatically identify locally aligned region pairs and employ a Bubble Sets--based visualization to highlight their structural properties (\textbf{R2}; \cref{sec:alignment_detection}). Each alignment pair is further categorized into canonical alignment patterns, with glyph-based querying enabling efficient exploration and comparison (\textbf{R3}; \cref{sec:alignment_detection}). 
Finally, we introduce a membrane-inspired visualization that supports multi-scale inspection of fine-grained alignment structures while reducing visual clutter (\textbf{R4}; \cref{sec:multiscale_alignment}).

\subsection{Global Alignment of Mapper Graphs}
\label{sec:alignment_layout}

Our goal is to compute globally aligned layouts for the two mapper graphs such that (1) each graph maintains a clear and readable structure, and (2) corresponding regions across graphs, determined by shared underlying data points, are spatially aligned. To achieve this, we propose a joint force-directed optimization framework that simultaneously computes the layouts of both graphs.

Specifically, we extend the standard force-directed layout formulation (see \cref{sec:background_fd}) by introducing \textit{cross-graph alignment forces}. Let
\[
\mathbf{p} = \{p_v \in \mathbb{R}^2 \mid v \in V_1 \cup V_2\}
\]
denote the 2D positions of all nodes. We define the overall energy function as
\[
\mathcal{L}(\mathbf{p}) =
\mathcal{L}_{1}^{\text{intra}}
+
\mathcal{L}_{2}^{\text{intra}}
+
\lambda \mathcal{L}_{\text{align}},
\]
where the intra-graph energy term $\mathcal{L}_{k}^{\text{intra}}$ for each graph $G_k$ consists of the standard attraction and repulsion terms described in~\cref{sec:background_fd}.

The cross-graph alignment term encourages nodes sharing common data points across graphs to be positioned closer together:
\[
\mathcal{L}_{\text{align}}
=
\sum_{(i,j) \in E_{\text{inter}}}
w_{ij} \, \|p_i - p_j\|^2,
\]
where $w_{ij}$ denotes the Jaccard similarity between the sets of data points associated with nodes $i$ and $j$:
\[
w_{ij}
=
\frac{|S(i) \cap S(j)|}{|S(i) \cup S(j)|}.
\]

The parameter $\lambda$ controls the trade-off between preserving the internal structure of each graph and aligning corresponding regions across graphs. The energy function $\mathcal{L}(\mathbf{p})$ is minimized iteratively by updating node positions according to the induced forces until convergence.

\para{Interactive visualization.}
After optimization, the two graphs are separated and displayed side by side. Users can interactively adjust $\lambda$ to explore layouts ranging from independent to strongly aligned configurations. We also considered an alternative strategy that fixes one graph while positioning the other relative to it; however, this introduces asymmetry and often leads to suboptimal alignment results.

\begin{figure}[!t]
\centering
\includegraphics[width=\linewidth]{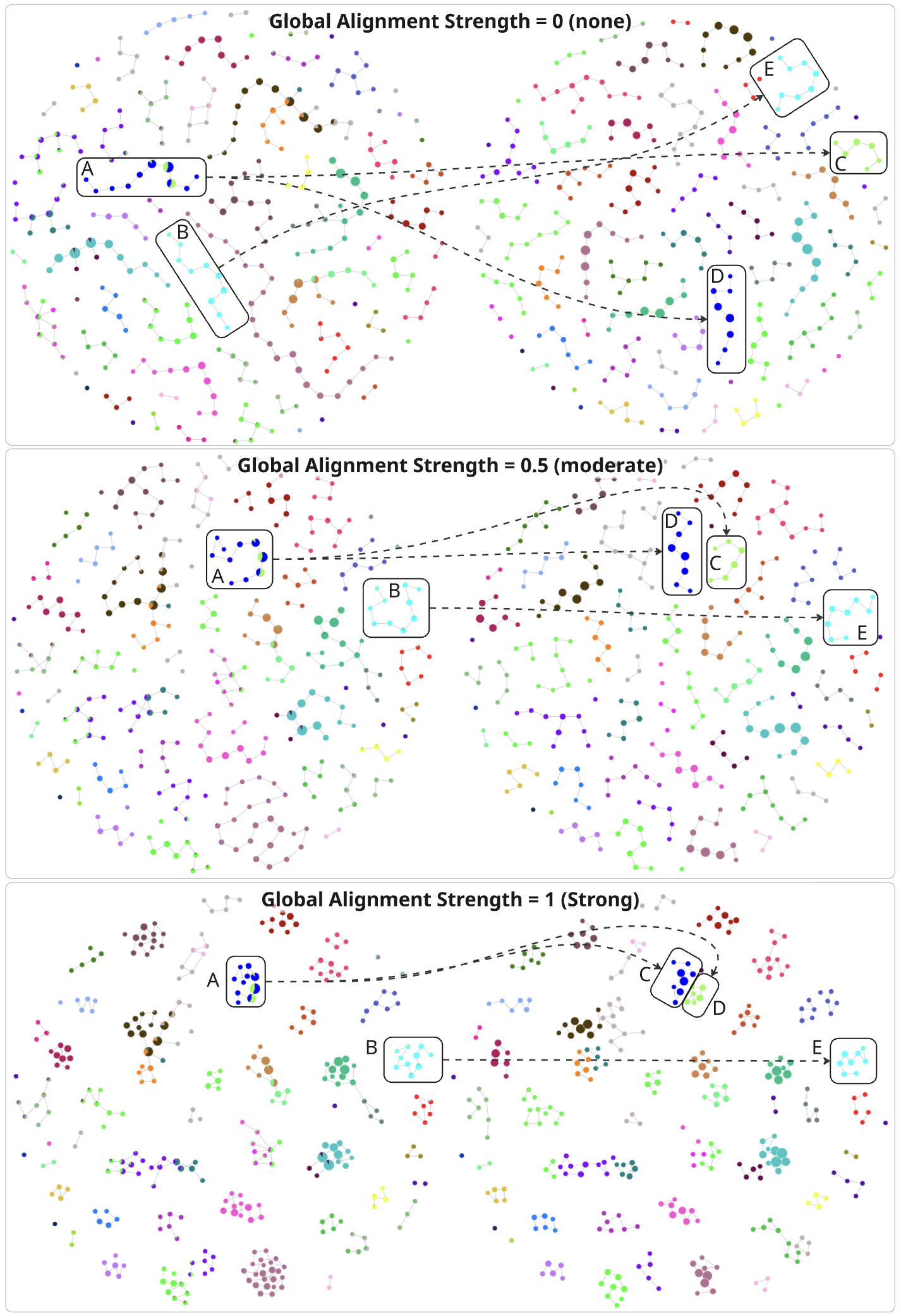}
\vspace{-4mm}
\caption{Mapper graphs of the final layer of BERT-base after fine-tuning for 2 epochs (left) and 6 epochs (right), illustrating the effect of global alignment with strengths 0, 0.5, and 1 (top to bottom). Highlighted correspondences show component A (left) aligning with components C and D (right), and component B (left) aligning with component E (right).}
\label{fig:global-alignment}
\vspace{-4mm}
\end{figure}

\cref{fig:global-alignment} shows three alignment results (top to bottom) obtained with different $\lambda$ values on mapper graphs constructed from word representations at different training stages (see~\cref{sec:LLM-case}). Approximate node correspondences can be inferred from node colors, which encode semantic categories.
When $\lambda = 0$, the two mapper graphs are laid out independently, causing corresponding regions to appear in unrelated spatial locations. With moderate alignment ($\lambda=0.5$), corresponding regions occupy similar relative positions within each graph, while stronger alignment ($\lambda=1$) further pulls matched nodes together to emphasize finer-grained structural correspondence.

\subsection{Automatic Local Alignment Discovery and Profiling} 
\label{sec:alignment_detection}

To enable local correspondence analysis, we automatically identify \textit{local alignment pairs} between the two mapper graphs. Each pair consists of one well-connected subgraph from each mapper that represents a coherent local region and shares some data points with its counterpart.

\para{Alignment-aware clustering formulation.}
We formulate local alignment detection as an \textit{alignment-aware clustering} problem on the joint graph. 
Graph clustering~\cite{aggarwal2010survey} partitions nodes into groups such that nodes within each group are densely connected, while connections between groups are relatively weak. 
Clustering on the joint graph directly achieves our goal: intra-graph edges preserve local structure within each mapper graph, and inter-graph edges link corresponding regions across graphs. Each cluster thus represents a local alignment pair.

\para{Spectral clustering.}
We perform alignment-aware clustering using spectral clustering~\cite{von2007tutorial}, as it effectively captures global graph connectivity while producing balanced clusters suitable for alignment detection. Edges within the first graph ($E_1$), the second graph ($E_2$), and across graphs ($E_{\text{inter}}$) are weighted by $\alpha$, $\beta$, and $\gamma$, respectively. By default, all weights are set to 1, treating all edges equally so that clustering is driven primarily by the overall graph structure. 
The number of clusters $k$ is determined from the eigenspectrum using the first elbow point. Full implementation details are provided in the supplement.

\para{Bubble Sets-based characterization.}
To highlight local alignment quality and location, we use a Bubble Sets~\cite{collins2009bubble} representation for each alignment pair, enclosing the corresponding regions in both mapper graphs.
We encodes two key properties for each bubble.
\begin{enumerate}[leftmargin=*,noitemsep]
\item \textit{Content alignment}: Measured by the Jaccard similarity between the data points contained in the corresponding subgraphs, reflecting the similarity of their underlying data content.
\item \textit{Structural coherence}: Measured by the silhouette score~\cite{rousseeuw1987silhouettes}, a standard clustering-quality metric (see supplement for details), which captures the extent to which nodes are well connected within each subgraph while remaining well separated from other clusters.
\end{enumerate} 
These properties are mapped to the visual channels of each bubble (see \cref{fig:BERT-epoch-compare}). Shading opacity encodes Jaccard similarity, where darker bubbles indicate greater content overlap between corresponding subgraphs. Bubble waviness encodes silhouette score, with smoother contours indicating stronger structural coherence. Following Görtler et al.~\cite{gortler2017bubble}, we adopt waviness to provide clear visual contrast between alignment qualities. 
This visual design enables users to quickly identify alignment pairs and compare their alignment quality across graphs.

\para{Interaction.}
Users can interactively adjust the edge weights $\alpha$, $\beta$, and $\gamma$ to explore alternative clustering results and alignment patterns (see~\cref{fig:BERT-epoch-compare,fig:local-alignment-compare}), as well as control the number of clusters.

We emphasize that the local alignment results are not intended to produce a single optimal solution; rather, they serve as an entry point for exploratory analysis of local correspondences. Although the Bubble Sets visualization encodes several key alignment properties, the framework is flexible and can incorporate additional metrics to characterize other aspects of alignment quality.

\subsection{Alignment Motif Query} 
\label{sec:alignment_Motif} 

To interpret each local alignment pair between two mapper graphs, we characterize their correspondences into high-level patterns inspired by TDA concepts such as feature appearance, persistence, splitting, and merging~\cite{carlsson2009topology}.  
In our context, we define a \textit{feature} as a coherent semantic region represented by a connected mapper subgraph. 

\begin{figure}[!ht]
\centering
\vspace{-0.5em}
\includegraphics[width=\linewidth]{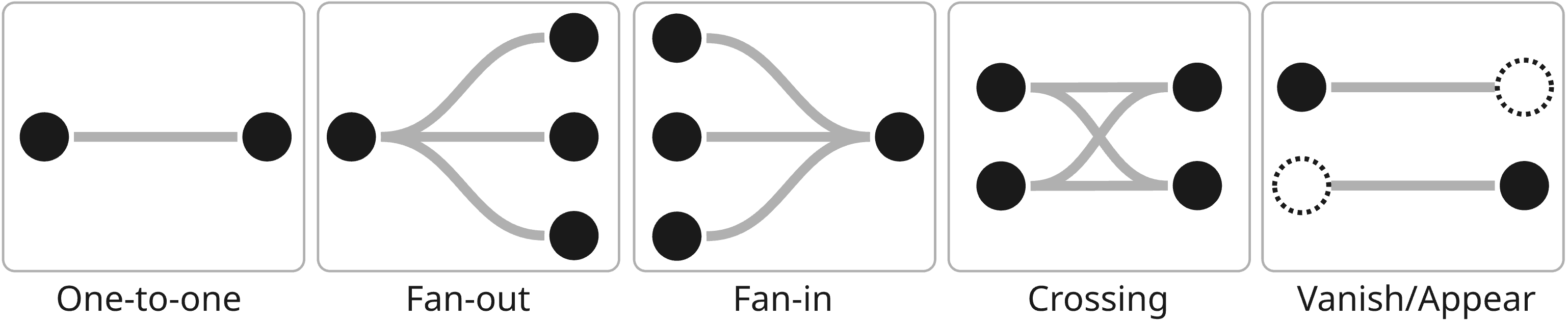}
\vspace{-2em}
\caption{An illustration of five local alignment patterns.}
\label{fig:alignment_patterns}
\vspace{-1em}
\end{figure}

Each alignment pair consists of two subgraphs, with correspondences determined by shared points connected via inter-graph edges.
We treat each component in a subgraph as an individual feature.
Using the inter-graph edges between these features, we categorize local alignment into five patterns, shown in~\cref{fig:alignment_patterns}.

\begin{itemize}[leftmargin=*,noitemsep]
    \item \textbf{One-to-one:} A single component in one mapper corresponds primarily to a single component in the other, indicating a consistent feature representation across models.
    
    \item \textbf{Fan-out:} A single component corresponds to multiple components in the other mapper, suggesting that its associated data points are distributed across several regions, reflecting a more fine-grained, fragmented, or disentangled representation.
    
    \item \textbf{Fan-in:} Multiple components correspond to a single component in the other mapper, indicating that data points from distinct regions are merged into a more aggregated representation.
    
    \item \textbf{Crossing:} Multiple components on one side align with multiple components on the other in a partially overlapping manner, indicating complex feature reorganization across models.
    
    \item \textbf{Vanishing/Appearance:} A component appears in only one mapper, highlighting model-specific features that are not captured by the other.
\end{itemize}

\para{Interaction and motif querying.}  
To facilitate analysis, each local alignment pair is assigned a pattern label based on its component-level correspondences. 
Users can select a pattern to filter and highlight all corresponding alignment pairs, enabling efficient motif-based querying and comparison across the two mapper graphs (see~\cref{fig:interface} B3).

\subsection{Fine-Grained Multiscale Alignment Analysis}
\label{sec:multiscale_alignment}

Given a local alignment pair, we support fine-grained inspection of correspondences between the two mapper subgraphs. Directly visualizing all nodes and inter-graph edges often leads to severe visual clutter, whereas collapsing each component into a single node oversimplifies the underlying structure. To balance these competing factors, we propose a multiscale visual encoding that progressively merges components into \emph{supernodes}, enabling users to explore alignment patterns at multiple levels of abstraction.

\para{Entropy-guided aggregation.} 
Our multiscale aggregation strategy is guided by the entropy of inter-edges between the two mapper subgraphs. Intuitively, high entropy indicates dispersed connections and less structured alignments, whereas low entropy reflects concentrated connections and more coherent alignment patterns. To reveal these structures, we iteratively merge pairs of connected nodes within each mapper subgraph. At each step, we select the merge that yields the greatest reduction in inter-edge entropy and continue until no further merge decreases the entropy.

A straightforward formulation defines the total inter-edge entropy as
\[
H_{\text{raw}} = - \sum_{(i,j) \in E_{\text{inter}}} p_{ij} \log p_{ij},
\quad 
p_{ij} = \frac{W_{ij}}{\sum_{(k,l) \in E_{\text{inter}}} W_{kl}},
\]
where $W_{ij}$ denotes the inter-edge weight (i.e., Jaccard similarity) between nodes $i$ and $j$. However, this formulation is inherently biased: merging nodes trivially reduces the number of inter-edges, causing $H_{\text{raw}}$ to decrease monotonically. As a result, the process ultimately collapses each component into a single supernode, potentially obscuring meaningful intermediate alignment structures.

To address this issue, we instead adopt a \emph{conditional-entropy} formulation. Each supernode $i$ is associated with a set of inter-edges connecting it to neighboring supernodes $\text{adj}(i)$ in the other mapper subgraph. We first compute a \textit{node-wise entropy} that measures the dispersion of its cross-graph connections:
\[
H_i = - \sum_{j \in \text{adj}(i)} p(j \mid i) \log p(j \mid i),
\quad
p(j \mid i) = \frac{W_{ij}}{\sum_{k \in \text{adj}(i)} W_{ik}}.
\]
We then define the \textit{global entropy} as a weighted sum over all participating supernodes:
\[
H_{\text{total}} = \sum_i w_i H_i,
\quad
w_i = \frac{\sum_{j \in \text{adj}(i)} W_{ij}}{\sum_u \sum_{v \in \text{adj}(u)} W_{uv}}.
\]
This formulation encourages merges that meaningfully reduce inter-edge uncertainty, allowing the aggregation process to terminate at intermediate stages that reveal coherent alignment structures, rather than always collapsing each component into a single supernode.

\para{Multiscale membrane visualization.}
Building on the aggregated structure, we introduce a membrane-inspired visualization in which the two mapper subgraphs are displayed as parallel layers connected by inter-edges. Each supernode is represented as an oval containing an outward-floating layout of its internal structure, inspired by the compartmentalized organization of biological membranes. This design supports analysis of cross-graph correspondences while preserving fine-grained local structural detail.

\para{Layout optimization.}
We employ a force-directed layout strategy for the membrane visualization. Each merged subgraph is first arranged using standard force-directed forces to preserve its internal structure. To improve alignment between the two layers, we introduce an additional constraint that vertically aligns corresponding merged subgraphs. Inter-graph edges are assigned a preferred length equal to the gap between the two layers, reducing excessively long or crossing connections.

Finally, the internal structure of each supernode is independently arranged using a local force-directed layout and displayed as a compact, outward-floating structure on its corresponding side.

\begin{figure}[!t]
\centering
\includegraphics[width=\linewidth]{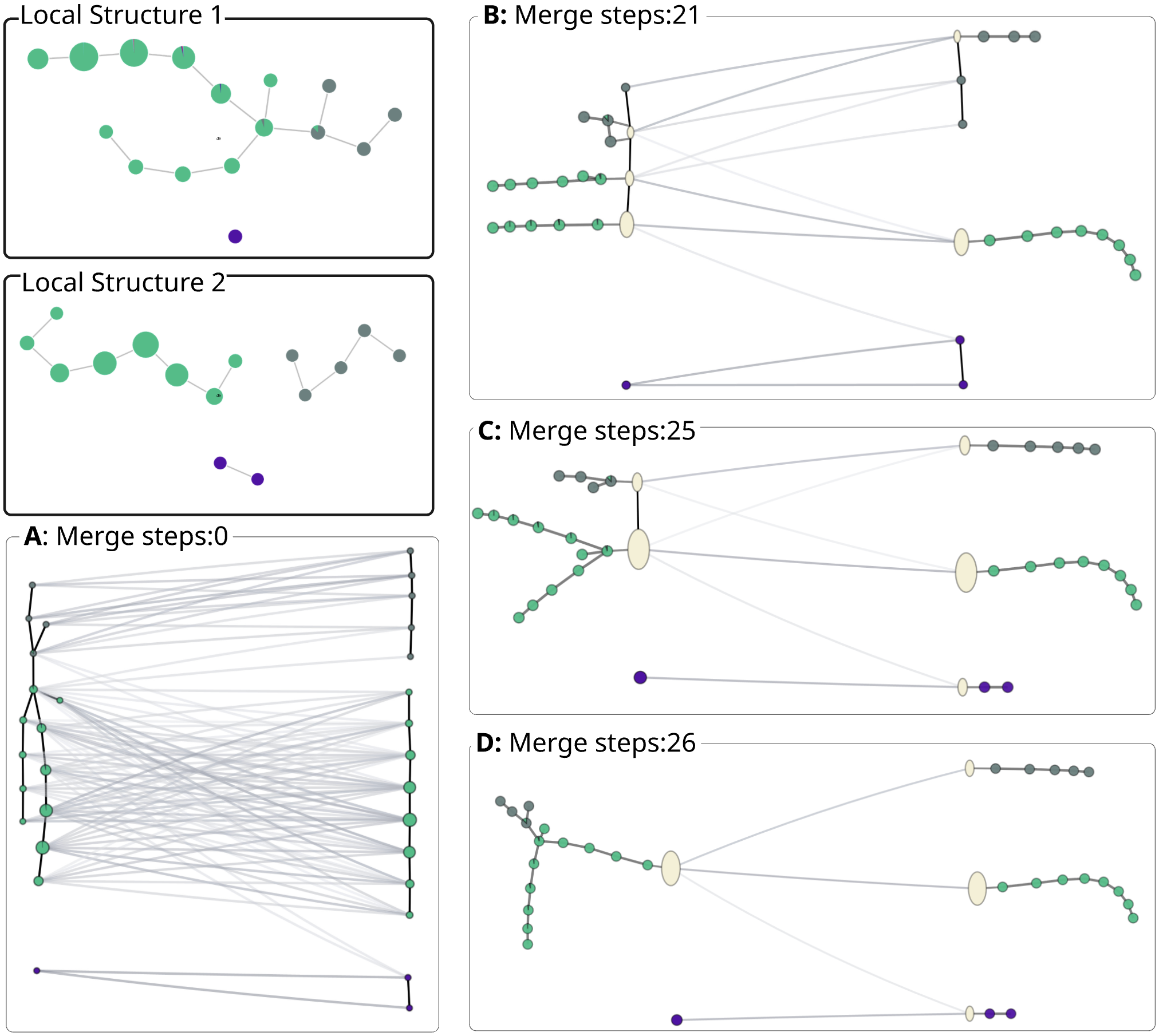}
\vspace{-2em}
\caption{A local alignment between structures 1 and 2 and their corresponding membrane views across different merging stages (A–D), with structure 1 shown on the left and structure 2 on the right. Panels C and D illustrate merging results obtained using the \textit{conditional-entropy} and \textit{raw-entropy} strategies, respectively (see~\cref{sec:multiscale_alignment}).}
\label{fig:merge-steps}
\vspace{-2em}
\end{figure}

\para{Interaction.}
Users can control the level of abstraction by adjusting the merge steps, interactively manipulate layout forces, drag nodes, and inspect the underlying data points. We illustrate the aggregation process in~\cref{fig:merge-steps} using a local alignment pair (structures 1 and 2) and their corresponding membrane views at different merge stages, with structure 1 shown on the left and structure 2 on the right.

In this example, structure 1 contains a component that partially mixes two data types (indicated by color), whereas these data types are separated in structure 2. Using the \textit{conditional-entropy} strategy (C), the mixed component is divided into two supernodes, allowing the mixed region to be revealed through the inter-edges (see details in~\cref{fig:BERT-epoch-compare} E). In contrast, the \textit{raw-entropy} strategy (D) merges each component into a single supernode, obscuring this alignment pattern.

\subsection{Workspace}

\begin{figure*}[!ht]
\centering
\includegraphics[width=\linewidth]{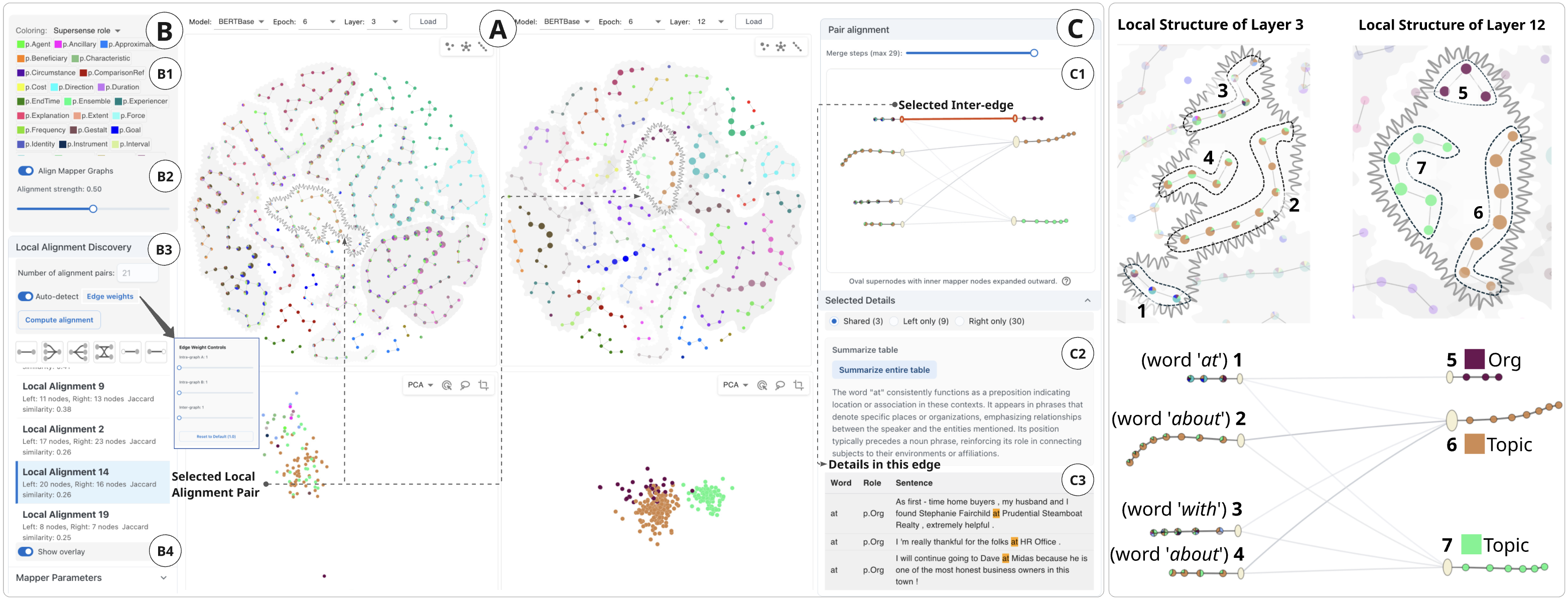}
\vspace{-6mm}
\caption{
{\tool} workspace comparing fine-tuned BERT-Base representations at layer 3 (left) and layer 12 (right), organized into three coordinated panels:
(A) \textbf{Overview Panel:} a 2$\times$2 layout showing two model representations, each consisting of a mapper graph (top) and a projection view (bottom).
(B) \textbf{Control Panel:} controls for coloring strategies (B1), global alignment strength (B2), local alignment discovery (B3), and alignment pattern filtering with Bubble Sets overlays (B4).
(C) \textbf{Detail Panel:} detailed information for a selected local alignment, including a membrane view (C1), LLM-generated summaries (C2), and a table of related items (C3).
In this example, a selected local alignment is highlighted across both mapper graphs and projection views, with additional details shown in the right panel.
}
\label{fig:interface}
\vspace{-2em}
\end{figure*} 

We integrate our visual designs (\cref{sec:alignment_layout,sec:alignment_detection,sec:alignment_Motif,sec:multiscale_alignment}) into a workspace prototype for interactive analysis of representation alignment, as shown in~\cref{fig:interface} (left).
The design follows the visualization mantra \textit{overview first, zoom and filter, then details on demand}~\cite{shneiderman2003eyes}, enabling exploration from global structures to local patterns (\textbf{T1}-\textbf{T3}). 
The system builds on established interaction and visualization techniques for mapper graphs~\cite{rathore2023topobert,zhou2021mapper,yan2025explainable} and projection views~\cite{boggust2022embedding}. 
We briefly introduce the interface; additional details are provided in the supplement.

\para{Overview Panel (A).} 
As shown in~\cref{fig:interface}, the central workspace adopts a 2$\times$2 layout. Each column corresponds to a model representation and contains a mapper graph (top) and a projection view (bottom). We include the projection views to help users interpret mapper structures, following recommendations from prior work~\cite{yan2025explainable,rathore2023topobert}.

\para{Control Panel (B).}
View B1 allows users to configure the coloring strategy following Zhou et al.~\cite{zhou2021mapper}. For categorical attributes, each mapper node is rendered as a pie chart indicating the proportion of categories, whereas numerical attributes are encoded using averaged values. By default, mapper node size is proportional to the number of data points contained within the node. The slider in B2 controls the global alignment strength between the two mapper graph layouts.

View B3 supports automatic detection of local alignments. Users can specify the number of alignment pairs or enable automatic detection, and adjust edge weights to balance intra- and inter-graph structures. The detected alignments are listed and can be filtered according to alignment patterns using the corresponding glyphs (see~\cref{sec:alignment_Motif}).

Enabling ``Show overlay'' in B4 displays Bubble Sets on the mapper graphs, revealing the spatial extent of each local structure. For each Bubble Set, shading opacity encodes Jaccard similarity (darker indicates higher similarity), while boundary fuzziness encodes silhouette score (more diffuse boundaries indicate higher scores). Hovering over a Bubble Set highlights the corresponding local structures in both mapper graphs.

To inspect a specific alignment pair, users can either select it from the list or directly click a Bubble Set. The system then highlights the corresponding local structures in both mapper graphs, filters the associated data points in the projection views, and displays detailed information in panel C. Users can also manually define a local structure on one mapper graph by selecting nodes, connected components, or shortest paths between nodes~\cite{zhou2021mapper,yan2025explainable}. The system subsequently identifies the corresponding structure in the other mapper graph to form a user-defined alignment, satisfying \textbf{R2-1}.

\para{Detail Panel (C).}
The right panel presents detailed information for the selected alignment pair. The membrane view (C1) visualizes intersections between the two local structures through inter-edges. Users can adjust a slider to control the number of merge steps, enabling inspection of alignment relationships at different levels of abstraction. The layout can also be refined through direct manipulation or layout controls (e.g., vertical alignment adjustments).

The table view (C3) displays the underlying data items, grouped into shared items and items unique to each local structure. Selecting an inter-edge in C1 highlights the corresponding items in the table. We further integrate the explainer agents proposed by Yan et al.~\cite{yan2025explainable}, using GPT-4o to automatically summarize the textual content of selected items (C2), thereby helping users interpret the semantic relationships within each alignment.
\section{Case Studies}
\label{sec:case-studies}
To demonstrate the effectiveness of {\tool}, we present two use cases: word representation alignment in language models (\cref{sec:LLM-case}) and image–text multimodal alignment (\cref{sec:CLIP-case}).

\subsection{Language Model Representation Alignment} 
\label{sec:LLM-case}
In this use case, we analyze how the latent spaces of language models evolve during fine-tuning across training epochs, layers, and model architectures through contextualized word representations.

\para{Data processing.} 
We fine-tune two transformer-based language models, BERT-base~\cite{devlin2019bert} and BERT-Tiny~\cite{jiao2020tinybert}, on a word classification task using the STREUSLE dataset~\cite{schneider2015corpus}, which provides word-level semantic annotations in context. Each target word, primarily a preposition, is assigned one of 57 supersense roles, corresponding to coarse-grained semantic categories such as \textit{location} and \textit{time}. The dataset contains 4,872 annotated words.

Both models are trained for six epochs using the same training configuration (see supplement). We extract contextualized representations of the annotated words from every layer at each training epoch: BERT-base contains 12 layers with 768-dimensional representations, while BERT-Tiny contains 2 layers with 128-dimensional representations.

In the following analyses, we examine representation dynamics across training epochs, layers, and models by constructing mapper graphs from these representations. Following TopoBERT~\cite{rathore2023topobert}, we use the $\ell^2$-norm as the filter function for each mapper, with 50 intervals and 50\% overlap. Clustering is performed using DBSCAN with $\texttt{min\_samples}=3$, while $\varepsilon$ is selected using the elbow method.

\subsubsection{Representation Alignment Across Training Epochs}
We use {\tool} to compare final-layer embeddings of BERT-base fine-tuned for 2 versus 6 epochs. \cref{fig:global-alignment} presents the corresponding mapper graphs (2 epochs on the left; 6 epochs on the right) under different global alignment strengths.
At a global level, the representations from epoch 2 exhibit more components with mixed label compositions, whereas the epoch-6 representations show clearer separation between semantic categories. This observation suggests that fine-tuning progressively disentangles semantic categories within the representation space over the course of training.

\begin{figure}[!b]
    \centering
    \vspace{-1.8em}
    \includegraphics[width=\linewidth]{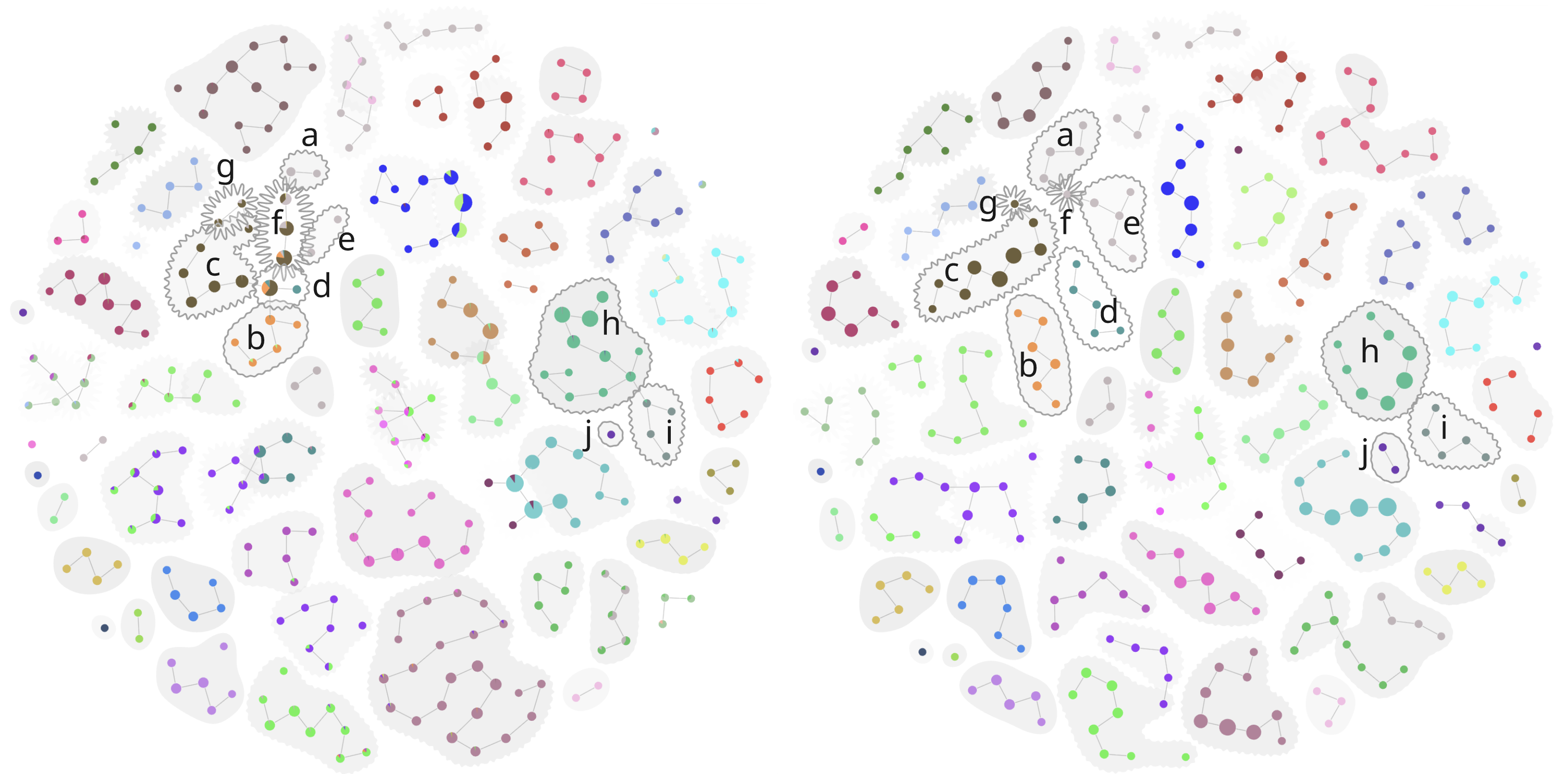}
    \vspace{-2em}
    \caption{Mapper graphs of the final layer of fine-tuned BERT-base at epochs 2 (left) and 6 (right). Each shaded Bubble Set represents a local alignment, with corresponding pairs positioned similarly across the two graphs. Alignments a–j are highlighted by boundary contours.}
    \label{fig:local-alignment-compare}
\end{figure} 

We apply the local alignment method to examine fine-grained structural correspondences between the two training stages. Using the default parameters (\cref{fig:local-alignment-compare}), we obtain 78 alignment pairs. Each local structure is enclosed by a Bubble Set, with corresponding alignments indicated through their relative placement across the two mapper graphs.
The resulting alignments largely preserve the structure of the right mapper while fragmenting several components in the left mapper. For example, a single component at epoch 2 is divided into seven local alignments (a--g), with a similar fragmentation pattern observed for (h, i). To better preserve the structure of the epoch-2 mapper, we increase the intra-graph edge weights for the epoch-2 graph during alignment optimization. 

As shown in~\cref{fig:BERT-epoch-compare} (left), this adjustment maintains the original epoch-2 structures more effectively; for instance, the previously separated alignments (a--g) are merged into a single local alignment F.
We then use these alignment results to analyze local structural patterns and compute prediction accuracy for instances contained within each aligned structure.

\para{One-to-one.} 
Most one-to-one alignments exhibit high prediction accuracy at both training stages. For example, pair A (\cref{fig:BERT-epoch-compare} A) aligns two components---one from each epoch---that both correspond to the label \textit{Whole} (e.g., ``\textit{[My]} hair is uneven...'') and achieve 100\% prediction accuracy. This suggests that the model learns this feature early during fine-tuning, indicating that the associated semantic pattern is relatively easy to acquire.

\para{Fan-in.}
In this pattern, multiple components at epoch 2 merge into a single component by epoch 6. For example, in pair~\cref{fig:BERT-epoch-compare} B, instances containing ``since,'' ``as,'' and ``after'' form one component, while instances containing ``for'' form another at epoch 2; by epoch 6, these representations merge into a single component. This transition suggests that fine-tuning progressively unifies representations belonging to the same semantic category.

\para{Fan-out.}
This pattern commonly occurs when a component containing mixed categories at epoch 2 splits into multiple cleaner components by epoch 6, often accompanied by improved prediction accuracy. In alignment D (\cref{fig:local-alignment-compare} D), words labeled as \textit{Theme} (e.g., ``it seems \textit{[like]} it's healthier too...'') and \textit{Manner} (e.g., ``I felt \textit{[like]} I was in heaven...'') are not well separated at epoch 2, resulting in 64.5\% prediction accuracy. By epoch 6, the two categories become fully separated, achieving 100\% accuracy. This progression suggests that the model continues to disentangle these semantic categories throughout fine-tuning.

Fan-out can also occur within a single category, where one component splits into multiple subcomponents. For example, in~\cref{fig:BERT-epoch-compare} C, the \textit{Identity} cluster at epoch 2 divides into three components by epoch 6: one containing ``as'' and two containing ``of,'' all achieving 100\% prediction accuracy. Such subdivision within the same category during later stages of fine-tuning may indicate overfitting.

\para{Crossing.}
This pattern combines characteristics of both fan-in and fan-out behaviors. As shown in~\cref{fig:BERT-epoch-compare} E, the categories \textit{Time}, \textit{StartTime}, and \textit{Circumstance} are partially mixed within a single component at epoch 2, but become clearly separated into distinct components by epoch 6.

The membrane view further clarifies this transition. At epoch 2, the mixed component forms two supernodes that roughly correspond to \textit{Time} and \textit{StartTime}; by epoch 6, it separates into three supernodes. Edges 1--3 between these supernodes highlight category mixing: each edge corresponds to an instance whose label differs from the dominant category of the associated supernode in the left mapper. Specifically, edge 1 represents a \textit{StartTime} instance embedded within a predominantly \textit{Time} cluster, edge 2 shows the reverse case, and edge 3 corresponds to a \textit{Circumstance} instance embedded within a predominantly \textit{Time} cluster. These boundary cases reveal ambiguities in the representation space that are progressively resolved through fine-tuning.

\subsubsection{Representation Alignment Between Two Layers} 
\label{sec:layer-compare}

\begin{figure}[!b]
    \centering
    \vspace{-1.5em}
    \includegraphics[width=\linewidth]{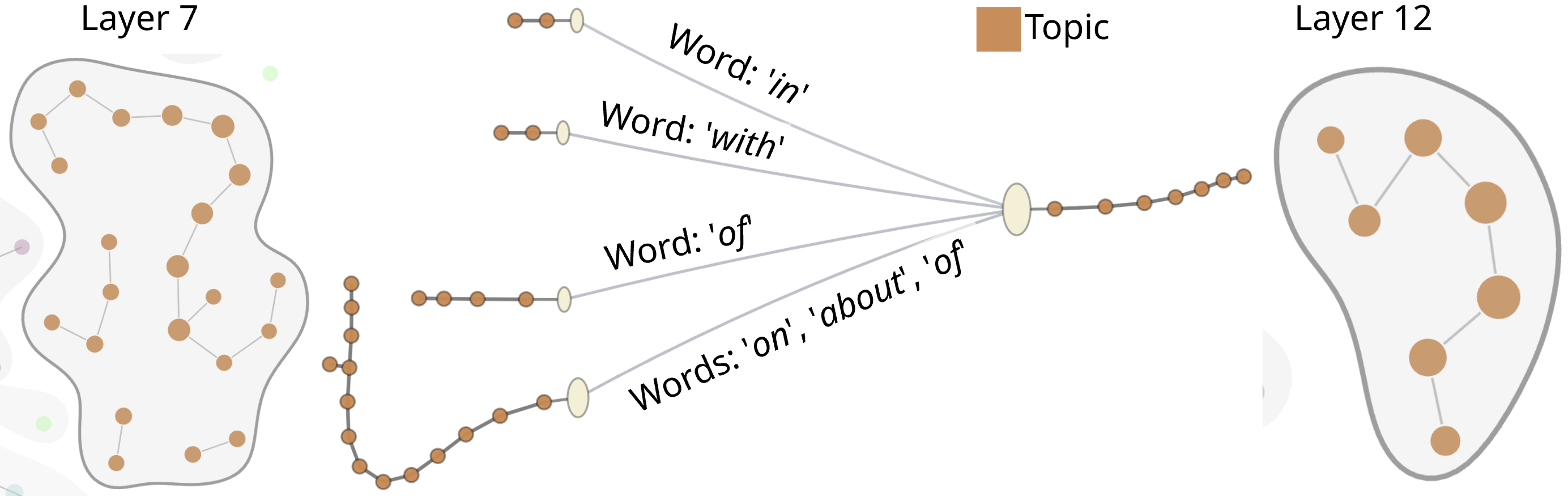}
    \vspace{-2em}
    \caption{A local alignment example exhibiting a fan-in pattern between layers 7 and 12 of the fine-tuned BERT-Base model.}
    \label{fig:Layer-7-12-compare}
\end{figure} 

We analyze how representations evolve across layers in BERT. Prior studies~\cite{rogers2020primer} suggest that early layers (1--3) primarily encode lexical information, whereas middle layers (6--8) capture richer semantic information.

\para{Layer 3 vs. Layer 12.}
As shown in~\cref{fig:interface} (left), representations in layer 3 are organized primarily by surface form: each component groups instances of the same word while largely ignoring semantic categories. In contrast, layer 12 exhibits clear category-based separation, indicating that early-layer representations are dominated by lexical similarity, whereas later layers become increasingly semantically organized.

Using local alignment, we identify specific instances of this transition. For example, in~\cref{fig:interface} (right), we observe a crossing pattern in which multiple components (supernodes 1--4) in layer 3---corresponding to the same word but different semantic categories---split and reorganize into distinct components (supernodes 5--7) in layer 12 that consistently reflect semantic categories. This pattern illustrates the progression from lexically driven groupings in early layers to task-aligned semantic structures in later layers.

\para{Layer 7 vs. Layer 12.}
Unlike layer 3, layer 7 already exhibits substantial semantic organization, with most components corresponding primarily to a single category. However, many categories remain distributed across multiple components, reflecting finer-grained semantic distinctions beyond the Supersense role labels.

As shown in the local alignment analysis (\cref{fig:Layer-7-12-compare}), words sharing the same label are divided across several clusters in layer 7, but consolidate into more unified category-specific structures by layer 12. This transition suggests a shift from fine-grained semantic representations in intermediate layers toward more task-aligned representations in the final layer.

\subsubsection{Representation Alignment Between Two Models}

\begin{figure}[!ht]
    \centering
    \includegraphics[width=\linewidth]{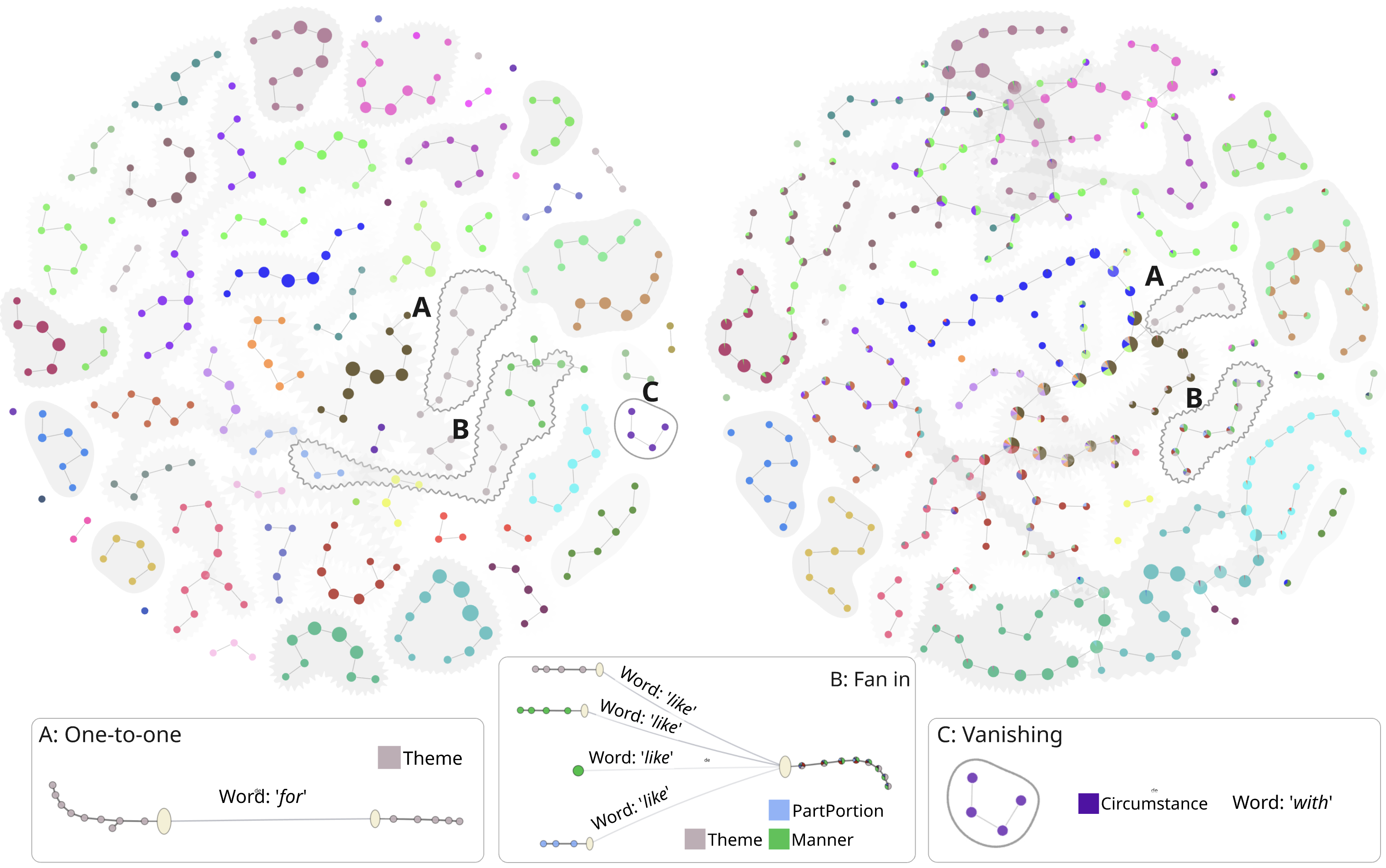}
    \vspace{-2em}
    \caption{
    Mapper graphs of the final layers of BERT-base (left) and BERT-Tiny (right). Local alignments are visualized using Bubble Sets, with alignments A–C highlighted by boundary contours. Corresponding membrane views are shown below.
    }
    \label{fig:BERT-model-compare}
    \vspace{-2em}
\end{figure} 

\cref{fig:BERT-model-compare} presents the aligned mapper graphs of BERT-Base and BERT-Tiny after six epochs of fine-tuning. At a global level, BERT-Tiny exhibits more entangled label structures than BERT-Base, reflecting its comparatively limited representational capacity due to its smaller model size. Using the local alignment method, we further analyze the best- and worst-performing local structures in BERT-Tiny.

The one-to-one alignment in \cref{fig:BERT-model-compare} A corresponds to the best-performing case, where both subgraphs achieve 100\% prediction accuracy for the \textit{Theme} category. The BERT-Tiny subgraph contains only the word ``for,'' suggesting that this semantic feature can be effectively learned even by the smaller model.

In contrast, the worst-performing alignment (\cref{fig:BERT-model-compare} B) reveals a substantial performance gap: BERT-Base achieves 100\% accuracy, whereas BERT-Tiny reaches only 58\%. In this case, the BERT-Tiny representation mixes three semantic categories associated with the word ``like,'' while BERT-Base clearly separates them, indicating that these distinctions are difficult for BERT-Tiny to learn.

We also observe local structures that appear only in BERT-Base, such as the \textit{Theme} cluster associated with the word ``with'' (\cref{fig:BERT-model-compare} C), suggesting that BERT-Tiny has not yet acquired this feature representation. These findings indicate that fine-tuning strategies for smaller models may benefit from emphasizing semantically difficult features, rather than applying uniform training across all categories.

\subsection{Multimodal Representation Alignment} 
\label{sec:CLIP-case}

\begin{figure}[!b]
    \centering
    \vspace{-1.5em}
    \includegraphics[width=\linewidth]{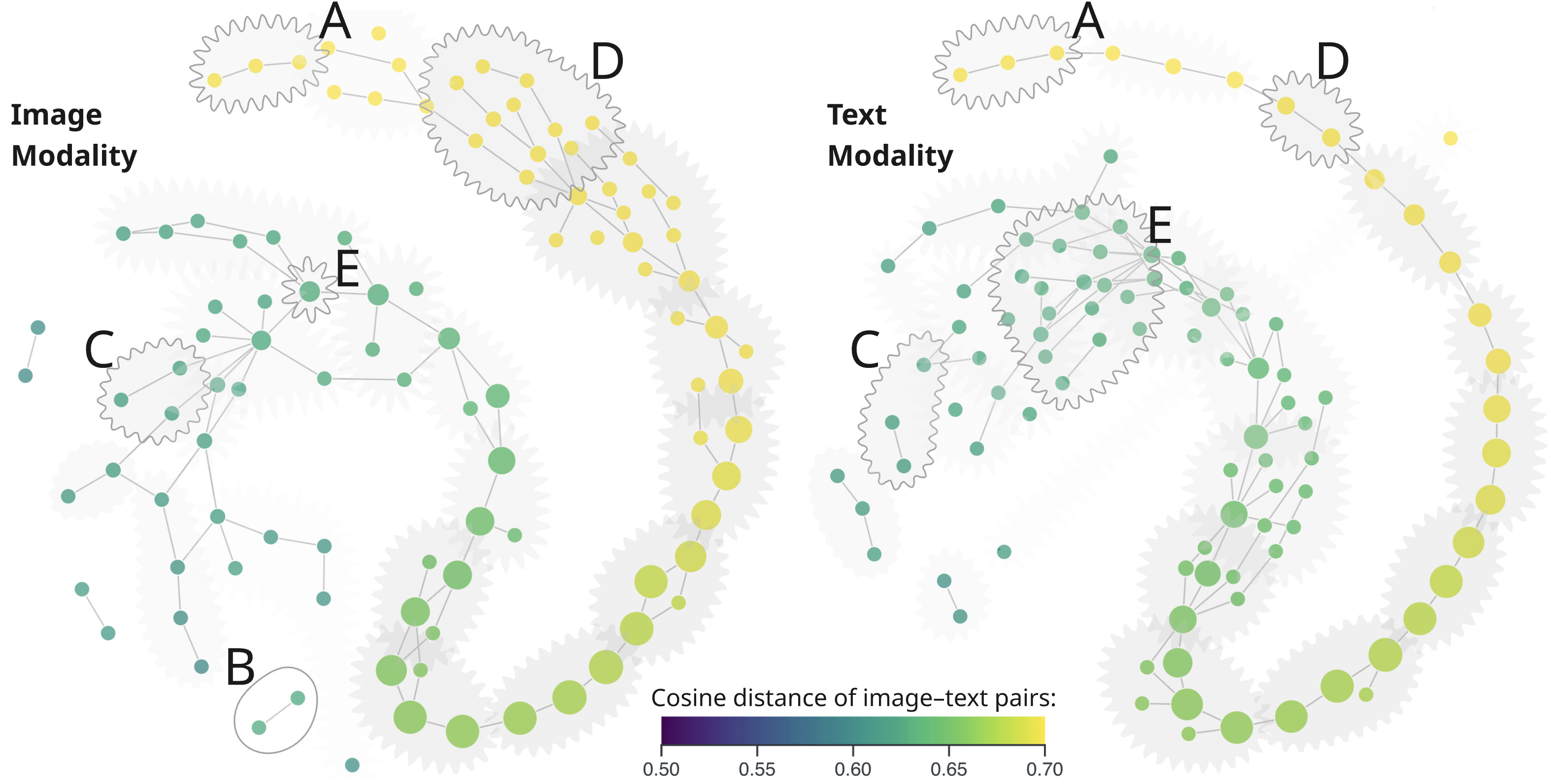}
    \vspace{-2em}
    \caption{Mapper graphs of the CLIP image modality (left) and caption modality (right). Nodes are colored by the average image–caption cosine distance (yellow indicates higher distance; blue indicates lower distance). Local alignments are visualized using Bubble Sets, with representative patterns (A–E) highlighted by boundary contours. Detailed views of these alignments are provided in~\cref{fig:CLIP-alignment-ABC,fig:CLIP-alignment-DE}.}
    \label{fig:CLIP-alignment-global}
\end{figure} 

We apply our approach to analyze alignment between image and text representations produced by CLIP~\cite{radford2021learning}, which maps both modalities into a shared embedding space and aligns image--caption pairs through contrastive learning.

\para{Data processing.} 
We use the MS COCO~\cite{lin2014microsoft} validation set, which contains 5,000 images, each paired with five captions and object annotations. Image and caption features are extracted using CLIP. For each image, we select the caption with the highest cosine similarity, resulting in 5,000 image--caption pairs.

For each modality, we construct a mapper graph using the image--caption cosine distance as the filter function (50 intervals with 50\% overlap). Clustering is performed using DBSCAN with $\texttt{min\_samples}=3$, while $\varepsilon$ is selected using the elbow method.

\para{Global alignment.}  
\cref{fig:CLIP-alignment-global} presents the globally aligned mapper graphs for the two modalities, where nodes are colored by the average image--caption cosine distance (yellow indicates high distance; blue indicates low distance). Both modalities exhibit a chain-like topology along the filter function, with finer-grained structures emerging in low-distance regions, suggesting more discriminative representations in these areas.

\para{Local alignment.}
Although the two modalities share similar global structures, they exhibit notable local differences. Using our alignment discovery method, we identify 24 local alignments, including 8 one-to-one, 5 fan-in, 6 fan-out, and 3 vanishing patterns, visualized using Bubble Sets in~\cref{fig:CLIP-alignment-global}.

\begin{figure}[!t]
    \centering
    \includegraphics[width=\linewidth]{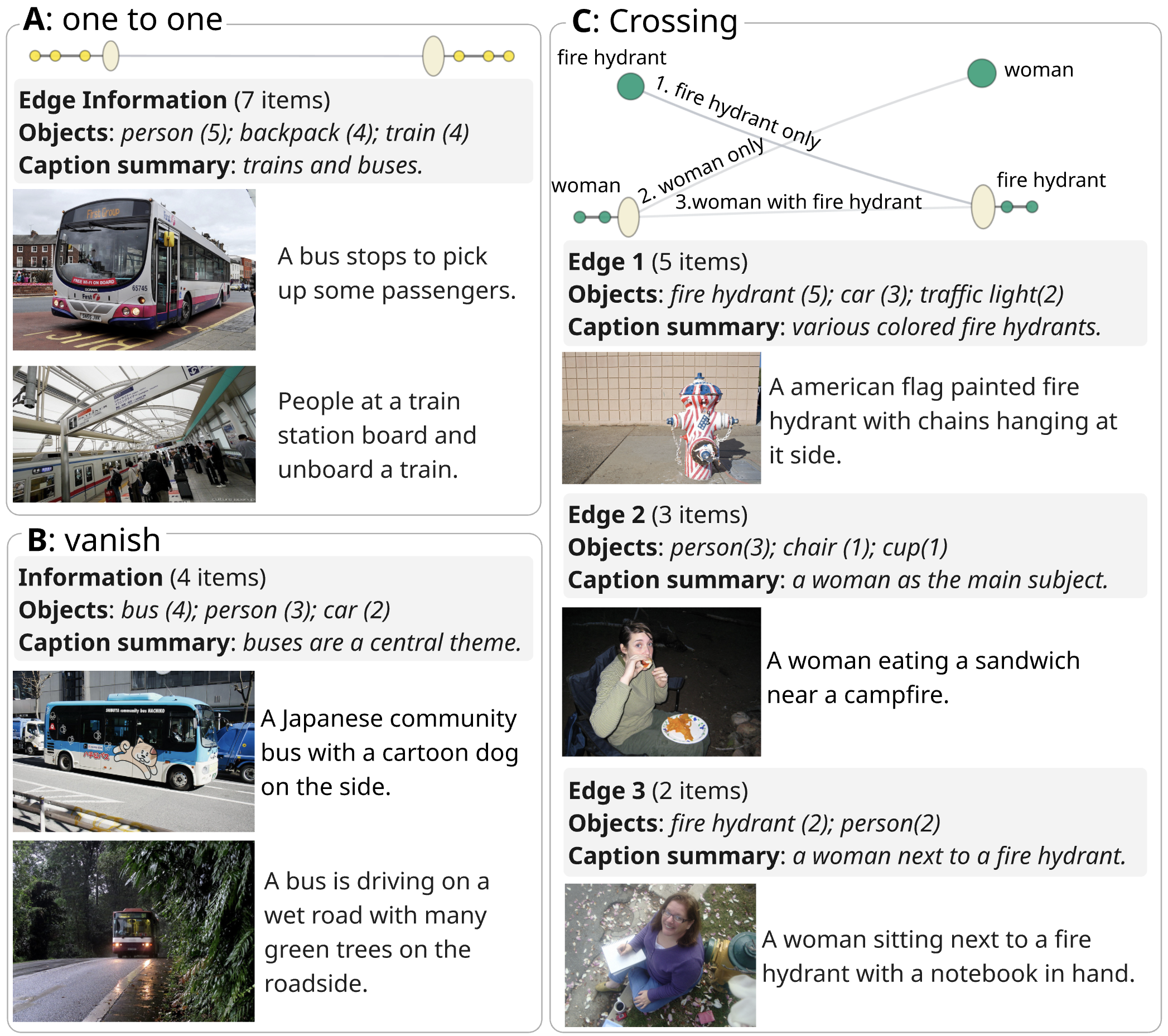}
    \vspace{-2em}
    \caption{Details of local alignments A, B, and C. For each alignment, we present a membrane visualization with image structures on the left and caption structures on the right, along with cross-modality edge information and an LLM-generated summary of the associated captions.}
    \label{fig:CLIP-alignment-ABC}
\vspace{-2em}
\end{figure}

\begin{figure}[!t]
    \centering
    \includegraphics[width=\linewidth]{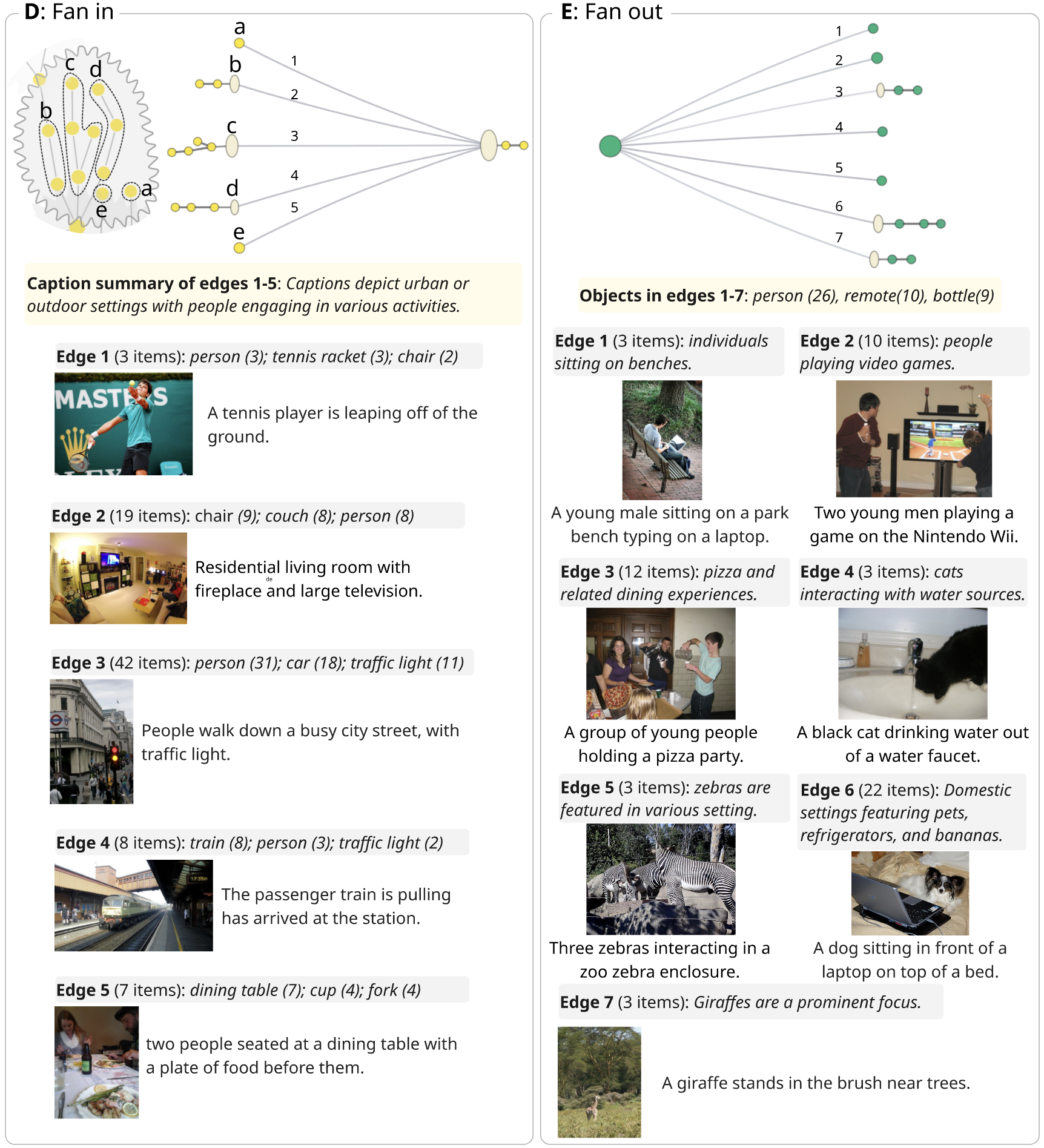}
    \vspace{-2em}
    \caption{Details of local alignments D and E. For each alignment, we present the membrane visualization (image modality on the left, caption modality on the right), along with cross-modality edge information and an LLM-generated caption summary. For alignment D, we additionally show the correspondence between supernodes and their underlying subgraphs.}
    \label{fig:CLIP-alignment-DE}
    \vspace{-2em}
\end{figure}

We further highlight five representative alignment patterns (\textit{A}--\textit{E}) in~\cref{fig:CLIP-alignment-global}, with their corresponding membrane views shown in~\cref{fig:CLIP-alignment-ABC,fig:CLIP-alignment-DE}. In each membrane view, the image modality is displayed on the left and the caption modality on the right.

For each cross-modality edge, we present the number of associated items, the frequencies of specific objects appearing in the images (top three most frequent categories derived from image metadata), and an LLM-generated summary of the associated captions.

\para{A: One-to-one.} 
As shown in~\cref{fig:CLIP-alignment-ABC} A, although alignment \textit{A} lies in a high-distance region, the image and caption nodes still share coarse semantic content, such as buses and trains. The relatively large image--caption distance may arise from low-quality captions that hallucinate unsupported details (e.g., ``to pick up some passengers'') or oversimplify visually complex scenes.

\para{B: Vanishing.} 
Component \textit{B} (\cref{fig:CLIP-alignment-ABC} B) appears in a low-distance region and exists only in the image mapper. The corresponding images depict bus-related scenes, while the captions provide fine-grained details, such as specific bus features and contextual descriptions. Although these details keep the image--caption pairs semantically close, they fragment the caption representations into multiple structures, causing the corresponding image structure to vanish in the alignment.

\para{C: Crossing.} 
Alignment \textit{C} (\cref{fig:CLIP-alignment-ABC} C) exhibits a crossing pattern caused by different concept groupings across modalities. The samples involve three related concepts: \textit{woman}, \textit{fire hydrant}, and \textit{woman with fire hydrant}. In the image modality, \textit{woman} and \textit{woman with fire hydrant} are grouped together and separated from \textit{fire hydrant only}. In contrast, the caption modality groups \textit{fire hydrant} and \textit{woman with fire hydrant} together while separating them from \textit{woman only}. The crossing edges therefore reveal how the two modalities partition and merge concepts differently, suggesting that they emphasize different semantic aspects of the same scenes.

\para{D: Fan-in.} 
Alignment \textit{D} (\cref{fig:CLIP-alignment-DE} D) occurs in a high-distance region, where the image modality exhibits more fine-grained structure than the caption modality. Along the cross-modality edges, the images form distinct scene groups, including tennis players, indoor rooms, trains, and dining tables, which all map to a single caption supernode. The associated captions are summarized by the LLM as ``urban or outdoor settings with people engaging in various activities.'' This pattern suggests that visually diverse scenes are described using relatively coarse textual descriptions, leading to larger image--caption distances.

\para{E: Fan-out.}
Alignment \textit{E} (\cref{fig:CLIP-alignment-DE} E) appears in a low-distance region. A single image node containing scenes of people and animals corresponds to multiple semantically specific caption groups, including ``person sitting on a bench,'' ``people playing video games,'' ``zebras,'' and ``cats interacting with water.'' This pattern suggests that caption representations capture finer semantic distinctions while remaining closely aligned with the image representations.

Together, these alignment patterns demonstrate that CLIP representations exhibit modality-specific differences in semantic granularity and emphasis.

\section{Expert Feedback}
\label{sec:evaluation}

To evaluate the effectiveness of {\tool}, we conducted a preliminary expert study with five participants (E1–E5): four PhD students and one postdoc in deep learning domain (e.g., computer vision, language models, and probabilistic learning), with 3.5–5 years of experience. 
All were familiar with model representations but not mapper graphs. 
Participants explored datasets based on their expertise: E1–E4 examined the CLIP case (\cref{sec:CLIP-case}), and E5 analyzed the BERT case (\cref{sec:LLM-case}). 
Each session included a 15-minute tutorial, 30 minutes of think-aloud exploration, and a 15-minute semi-structured interview. 
All sessions were recorded.
We summary key insights below.

\para{Global alignment.}
All participants found the global alignment layout effective for identifying correspondences between the two mapper graphs, avoiding manual inspection of individual nodes.
E1 expressed interest in quantitative measures of alignment, while E3 initially misinterpreted node positions as indicating cluster proximity, suggesting a need for clearer guidance. 
E4 commented that \textit{``the smooth animation for alignment strength gives a good intuition of what is happening.''}

\para{Alignment discovery and categorization.}
Participants found automatic alignment discovery helpful for quickly locating corresponding regions and reducing exploration effort. 
E4 noted that \textit{``the patterns seem pretty clear to me—I like the selection of different motifs,''} while E5 commented that \textit{``the categorization is very intuitive and helpful for understanding high-level patterns.''} 
E1 further expressed interest in better understanding the underlying computation.

\para{Bubble Sets visualization.}
The Bubble Sets design was generally well received. 
E1 noted that it \textit{``helps to select structures based on the fuzzy boundary.''} 
E4 suggested adding semantic annotations (e.g., captions) directly onto Bubble Sets to streamline exploration, while E5 felt the overlay could become visually overwhelming in dense regions.

\para{Membrane view.}
Feedback on the membrane view was mixed. 
E3 found it intuitive for simplifying inter-graph connections and liked the distinct visual encoding of supernodes. 
E1 was confused by supernode semantics, interpreting connections as linking to all constituent nodes
E4 found inter-edges hard to interpret without context and suggested semantic summaries (e.g., captions).  
E5 proposed alternative encodings (e.g., circular containers) to better show internal structure.

\para{Interface.}
Participants found the system easy to use. 
E2 proposed tighter coordination between the projection and mapper views (e.g., selecting subsets in the projection to reveal corresponding mapper nodes). E4 found some interactions, such as membrane layout sliders, less useful and suggested more task-driven exploration (e.g., querying ``people in motion'').

\para{Potential applications.}
Participants expressed strong interest to apply {\tool} to their own research. E1 suggested comparing models before and after fine-tuning to diagnose category-specific performance drops. E3 proposed assessing dataset labeling quality by comparing image and label modalities. E5 highlighted its use in detecting overfitting via overly fine-grained structures.

\section{Limitations and Discussions}
\label{sec:limitations}

\para{Local alignment discovery.}
Our automatic local alignment detection method leverages the structural properties of mapper graphs. Rather than seeking a single optimal alignment, the method is designed to surface potentially meaningful correspondences, recognizing that useful alignments often depend on user intent and analytical goals. Its primary purpose is therefore to help users quickly identify regions that may warrant further investigation.

Because the method operates at the cluster level, it can produce low-quality alignments when the underlying representation spaces differ substantially and lack clear node correspondences. To address this limitation, {\tool} allows users to manually select regions of interest and define custom local alignments. Future work will further support query-driven alignment, enabling users to retrieve mapper structures by querying specific words, concepts, or image classes, as motivated by E4.

\para{Automatic alignment interpretation and annotations.}
Interpreting a local alignment requires understanding both node semantics and the inter-graph relationships between subgraphs. Currently, we use LLMs to summarize textual content, but users must still manually compare semantic relationships across inter-edges and nodes, as well as inspect node- or subgraph-specific data points to derive deeper insights.

Future work will extend these explainer agents to automatically summarize and contrast alignment patterns, generate annotations, and support image-based summaries~\cite{yan2025vislix}, thereby improving semantic interpretation of local alignments.

\para{Multi-graph comparisons.}
At present, {\tool} supports pairwise comparison between mapper graphs. Future work will extend the framework to support multi-graph comparisons in two directions: (1) sequential comparisons across layers or training stages, and (2) comparisons across multiple independent models.

Sequential comparisons would enable users to track the evolution of representations over time or depth. For example, as observed in~\cref{sec:layer-compare}, early BERT layers cluster instances primarily by lexical identity, intermediate layers separate words according to semantic categories, and later layers group semantically related words across lexical forms, revealing the progression of feature organization during training.

\para{Expert feedback study.}
We conducted a preliminary study with a small group of participants to gather initial feedback on the design and utility of {\tool}. Given participants' limited prior experience with mapper graphs and the short duration of the sessions, the feedback primarily reflects initial impressions rather than long-term usage. Future work will involve extended studies with a larger participant pool to support more comprehensive evaluation and iterative refinement.

\para{Insights from representation alignment.}
Prior work~\cite{klabunde2025similarity} suggests that internal representation alignment does not necessarily correlate directly with model performance; for example, semantically ambiguous regions may still produce correct predictions (\cref{fig:BERT-epoch-compare} D). Nevertheless, performance metrics can help validate and contextualize observed alignment patterns~\cite{klabunde2025similarity}. For instance, mixed or entangled components are often associated with lower prediction accuracy (\cref{fig:BERT-epoch-compare} E).

More broadly, topological alignment reveals underlying semantic organization within representation spaces and can support downstream analyses such as diagnosing model weaknesses, studying representation evolution, and understanding how models organize knowledge.

\section{Conclusion}
\label{sec:conclusion}

In this work, we presented {\tool}, a topology-aware framework for comparing neural network representations through their structural organization. By leveraging mapper graphs, {\tool} supports multi-scale analysis of alignment between pairs of representations, ranging from globally aligned layouts to local correspondences and fine-grained structural patterns.

Our approach integrates joint layout optimization, alignment-aware clustering, and novel visual encodings to facilitate exploration of representation alignment. Through use cases involving language models and multimodal representations, we demonstrated that {\tool} reveals meaningful structural patterns across training stages, network layers, model architectures, and modalities. Preliminary expert feedback further suggests the effectiveness of the approach while highlighting opportunities for future refinement.

To the best of our knowledge, this work represents the first application of topological data analysis to representation alignment, and we hope it will inspire new directions for structure-aware analysis of neural representations.

\acknowledgments{This work was partially supported by grants from the U.S. National Science Foundation (projects IIS-2205418, and DMS-2134223) and the Swiss National Science Foundation (project
10003068).}

\bibliographystyle{abbrv-doi-hyperref-narrow}
\bibliography{refs-topoalign.bib}

@inproceedings{mehrer2018beware,
	author = {Mehrer, Johannes and Kriegeskorte, Nikolaus and Kietzmann, Tim},
	booktitle = {Conference on Cognitive Computational Neuroscience},
	doi = {10.32470/CCN.2018.1172-0},
	title = {Beware of the beginnings: intermediate and higherlevel representations in deep neural networks are strongly affected by weight initialization},
	year = {2018}}

@article{collins2009bubble,
	author = {Collins, Christopher and Penn, Gerald and Carpendale, Sheelagh},
	doi = {10.1109/TVCG.2009.122},
	journal = {IEEE Transactions on Visualization and Computer Graphics},
	number = {6},
	pages = {1009-1016},
	title = {{Bubble Sets}: Revealing Set Relations with Isocontours over Existing Visualizations},
	volume = {15},
	year = {2009}}

@inproceedings{klabunde2025resi,
	author = {Max Klabunde and Tassilo Wald and Tobias Schumacher and Klaus Maier-Hein and Markus Strohmaier and Florian Lemmerich},
	booktitle = {The Thirteenth International Conference on Learning Representations},
	title = {{ReSi}: A Comprehensive Benchmark for Representational Similarity Measures},
	url = {https://openreview.net/forum?id=PRvdO3nfFi},
	year = {2025}}

@article{atakishiyev2024explainable,
	author = {Atakishiyev, Shahin and Salameh, Mohammad and Yao, Hengshuai and Goebel, Randy},
	doi = {10.1109/ACCESS.2024.3431437},
	journal = {IEEE Access},
	pages = {101603-101625},
	title = {Explainable Artificial Intelligence for Autonomous Driving: A Comprehensive Overview and Field Guide for Future Research Directions},
	volume = {12},
	year = {2024}}

@inproceedings{arendt2020parallel,
	address = {New York, NY, USA},
	author = {Arendt, Dustin L. and Nur, Nasheen and Huang, Zhuanyi and Fair, Gabriel and Dou, Wenwen},
	booktitle = {Proceedings of the 25th International Conference on Intelligent User Interfaces},
	doi = {10.1145/3377325.3377514},
	pages = {259--274},
	publisher = {Association for Computing Machinery},
	title = {Parallel embeddings: a visualization technique for contrasting learned representations},
	year = {2020}}

@incollection{aggarwal2010survey,
	address = {Boston, MA},
	author = {Aggarwal, Charu C. and Wang, Haixun},
	booktitle = {Managing and Mining Graph Data},
	doi = {10.1007/978-1-4419-6045-0_9},
	editor = {Aggarwal, Charu C. and Wang, Haixun},
	pages = {275--301},
	publisher = {Springer US},
	title = {A Survey of Clustering Algorithms for Graph Data},
	year = {2010}}

@inproceedings{boggust2022embedding,
	address = {New York, NY, USA},
	author = {Boggust, Angie and Carter, Brandon and Satyanarayan, Arvind},
	booktitle = {Proceedings of the 27th International Conference on Intelligent User Interfaces},
	doi = {10.1145/3490099.3511122},
	pages = {746--766},
	publisher = {Association for Computing Machinery},
	title = {{Embedding Comparator}: Visualizing Differences in Global Structure and Local Neighborhoods via Small Multiples},
	year = {2022}}

@article{yan2025vislix,
	author = {Yan, Xinyuan and Xuan, Xiwei and Ono, Jorge Piazentin and Guo, Jiajing and Mohanty, Vikram and Kumar, Shekar Arvind and Gou, Liang and Wang, Bei and Ren, Liu},
	doi = {10.1111/cgf.70125},
	journal = {Computer Graphics Forum},
	number = {3},
	pages = {e70125},
	title = {{VISLIX}: An {XAI} Framework for Validating Vision Models with Slice Discovery and Analysis},
	volume = {44},
	year = {2025}}

@inproceedings{wu2025measuring,
	author = {Jialin Wu and Shreya Saha and Yiqing Bo and Meenakshi Khosla},
	booktitle = {UniReps: 3rd Edition of the Workshop on Unifying Representations in Neural Models},
	title = {Measuring the Measures: Discriminative Capacity of Representational Similarity Metrics Across Model Families},
	url = {https://openreview.net/forum?id=eELCEkN2Kg},
	year = {2025}}

@article{reddy2025towards,
	author = {Reddy, Chandan K and Shojaee, Parshin},
	doi = {10.1609/aaai.v39i27.35084},
	journal = {Proceedings of the AAAI Conference on Artificial Intelligence},
	number = {27},
	pages = {28601--28609},
	title = {Towards Scientific Discovery with Generative AI: Progress, Opportunities, and Challenges},
	volume = {39},
	year = {2025}}

@inproceedings{edamadaka2025universally,
	author = {Sathya Edamadaka and Soojung Yang and Rafael Gomez-Bombarelli},
	booktitle = {UniReps: 3rd Edition of the Workshop on Unifying Representations in Neural Models},
	doi = {10.48550/arXiv.2512.03750},
	title = {Universally Converging Representations of Matter Across Scientific Foundation Models},
	url = {https://openreview.net/forum?id=DTVW9v84yP},
	year = {2025}}

@article{rogers2020primer,
	author = {Rogers, Anna and Kovaleva, Olga and Rumshisky, Anna},
	doi = {10.1162/tacl_a_00349},
	journal = {Transactions of the association for computational linguistics},
	pages = {842--866},
	title = {A primer in BERTology: What we know about how BERT works},
	volume = {8},
	year = {2020}}

@article{cheong2020force,
	author = {Cheong, Se Hang and Si, Yain Whar},
	doi = {10.1177/1473871618821740},
	journal = {Information Visualization},
	month = {01},
	pages = {147387161882174},
	title = {Force-directed algorithms for schematic drawings and placement: A survey},
	volume = {19},
	year = {2019}}

@article{fruchterman1991graph,
	author = {Fruchterman, Thomas M. J. and Reingold, Edward M.},
	doi = {https://doi.org/10.1002/spe.4380211102},
	eprint = {https://onlinelibrary.wiley.com/doi/pdf/10.1002/spe.4380211102},
	journal = {Software: Practice and Experience},
	number = {11},
	pages = {1129-1164},
	title = {Graph drawing by force-directed placement},
	url = {https://onlinelibrary.wiley.com/doi/abs/10.1002/spe.4380211102},
	volume = {21},
	year = {1991}}

@article{kamada1989algorithm,
	author = {Tomihisa Kamada and Satoru Kawai},
	doi = {https://doi.org/10.1016/0020-0190(89)90102-6},
	issn = {0020-0190},
	journal = {Information Processing Letters},
	number = {1},
	pages = {7-15},
	title = {An algorithm for drawing general undirected graphs},
	url = {https://www.sciencedirect.com/science/article/pii/0020019089901026},
	volume = {31},
	year = {1989}}

@article{hu2005efficient,
	author = {Yifan Hu},
	journal = {The Mathematica journal},
	pages = {37-71},
	title = {Efficient, High-Quality Force-Directed Graph Drawing},
	url = {https://api.semanticscholar.org/CorpusID:14599587},
	volume = {10},
	year = {2006}}

@article{zhong2023force,
	author = {Zhong, Fahai and Xue, Mingliang and Zhang, Jian and Zhang, Fan and Ban, Rui and Deussen, Oliver and Wang, Yunhai},
	doi = {10.1109/TVCG.2023.3238821},
	journal = {IEEE Transactions on Visualization and Computer Graphics},
	number = {7},
	pages = {3650--3663},
	publisher = {IEEE},
	title = {Force-directed graph layouts revisited: a new force based on the t-distribution},
	volume = {30},
	year = {2023}}

@article{xue2025autofdp,
	author = {Xue, Mingliang and Wang, Yifan and Wang, Zhi and Zhu, Lifeng and Cui, Lizhen and Chen, Yueguo and Ding, Zhiyu and Deussen, Oliver and Wang, Yunhai},
	doi = {10.1109/TVCG.2025.3631659},
	journal = {IEEE Transactions on Visualization and Computer Graphics},
	number = {2},
	pages = {1554-1568},
	title = {{AutoFDP}: Automatic Force-Based Model Selection for Multicriteria Graph Drawing},
	volume = {32},
	year = {2026}}

@article{gortler2017bubble,
	author = {G{\"o}rtler, Jochen and Schulz, Christoph and Weiskopf, Daniel and Deussen, Oliver},
	doi = {10.1109/TVCG.2017.2743959},
	journal = {IEEE Transactions on Visualization and Computer Graphics},
	number = {1},
	pages = {719-728},
	title = {Bubble Treemaps for Uncertainty Visualization},
	volume = {24},
	year = {2018}}

@article{rousseeuw1987silhouettes,
	author = {Rousseeuw, Peter J},
	doi = {https://doi.org/10.1016/0377-0427(87)90125-7},
	journal = {Journal of computational and applied mathematics},
	pages = {53--65},
	publisher = {Elsevier},
	title = {Silhouettes: a graphical aid to the interpretation and validation of cluster analysis},
	volume = {20},
	year = {1987}}

@article{von2007tutorial,
	author = {Von Luxburg, Ulrike},
	doi = {10.1007/s11222-007-9033-z},
	journal = {Statistics and computing},
	number = {4},
	pages = {395--416},
	publisher = {Springer},
	title = {A tutorial on spectral clustering},
	volume = {17},
	year = {2007}}

@inproceedings{doppalapudi2022untangling,
	author = {Doppalapudi, Bhavana and Wang, Bei and Rosen, Paul},
	booktitle = {2022 Topological Data Analysis and Visualization (TopoInVis)},
	doi = {10.1109/TopoInVis57755.2022.00015},
	organization = {IEEE},
	pages = {81--91},
	title = {Untangling force-directed layouts using persistent homology},
	year = {2022}}

@article{purvine2023experimental,
	author = {Purvine, Emilie and Brown, Davis and Jefferson, Brett and Joslyn, Cliff and Praggastis, Brenda and Rathore, Archit and Shapiro, Madelyn and Wang, Bei and Zhou, Youjia},
	doi = {10.1609/aaai.v37i8.26134},
	journal = {Proceedings of the AAAI Conference on Artificial Intelligence},
	number = {8},
	pages = {9470-9479},
	title = {Experimental Observations of the Topology of Convolutional Neural Network Activations},
	volume = {37},
	year = {2023}}

@inproceedings{zhou2021mapper,
	author = {Zhou, Youjia and Chalapathi, Nithin and Rathore, Archit and Zhao, Yaodong and Wang, Bei},
	booktitle = {IEEE 14th Pacific Visualization Symposium},
	doi = {10.1109/PacificVis52677.2021.00021},
	pages = {101-110},
	title = {{Mapper Interactive}: A Scalable, Extendable, and Interactive Toolbox for the Visual Exploration of High-Dimensional Data},
	year = {2021}}

@inproceedings{zhou2023visualizing,
	author = {Zhou, Youjia and Zhou, Yi and Ding, Jie and Wang, Bei},
	booktitle = {Proceedings of 2nd Annual Workshop on Topology, Algebra, and Geometry in Machine Learning},
	editor = {Doster, Timothy and Emerson, Tegan and Kvinge, Henry and Miolane, Nina and Papillon, Mathilde and Rieck, Bastian and Sanborn, Sophia},
	pages = {134-145},
	publisher = {PMLR},
	series = {Proceedings of Machine Learning Research},
	title = {Visualizing and Analyzing the Topology of Neuron Activations in Deep Adversarial Training},
	volume = {221},
	year = {2023}}

@article{gleicher2011visual,
	author = {Gleicher, Michael and Albers, Danielle and Walker, Rick and Jusufi, Ilir and Hansen, Charles D and Roberts, Jonathan C},
	doi = {10.1177/1473871611416549},
	journal = {Information Visualization},
	number = {4},
	pages = {289--309},
	publisher = {SAGE Publications Sage UK: London, England},
	title = {Visual comparison for information visualization},
	volume = {10},
	year = {2011}}

@article{carlsson2009topology,
	author = {Carlsson, Gunnar},
	doi = {10.1090/S0273-0979-09-01249-X},
	journal = {Bulletin of The American Mathematical Society - BULL AMER MATH SOC},
	month = {04},
	pages = {255-308},
	title = {Topology and Data},
	volume = {46},
	year = {2009}}

@article{singh2025survey,
	author = {Singh, Sonali Uttam and Namin, Akbar Siami},
	doi = {https://doi.org/10.1016/j.nlp.2025.100128},
	journal = {Natural Language Processing Journal},
	pages = {100128},
	publisher = {Elsevier},
	title = {A survey on chatbots and large language models: Testing and evaluation techniques},
	volume = {10},
	year = {2025}}

@inproceedings{jones2022if,
	author = {Jones, Haydn T. and Springer, Jacob M. and Kenyon, Garrett T. and Moore, Juston S.},
	booktitle = {Proceedings of the Thirty-Eighth Conference on Uncertainty in Artificial Intelligence},
	editor = {Cussens, James and Zhang, Kun},
	month = {01--05 Aug},
	pages = {928--937},
	publisher = {PMLR},
	series = {Proceedings of Machine Learning Research},
	title = {If you've trained one you've trained them all: inter-architecture similarity increases with robustness},
	url = {https://proceedings.mlr.press/v180/jones22a.html},
	volume = {180},
	year = {2022}}

@article{kriegeskorte2008representational,
	author = {Kriegeskorte, Nikolaus and Mur, Marieke and Bandettini, Peter A},
	doi = {10.3389/neuro.06.004.2008},
	journal = {Frontiers in systems neuroscience},
	pages = {249},
	publisher = {Frontiers},
	title = {Representational similarity analysis-connecting the branches of systems neuroscience},
	volume = {2},
	year = {2008}}

@inproceedings{nanda2022measuring,
	author = {Nanda, Vedant and Speicher, Till and Kolling, Camila and Dickerson, John P and Gummadi, Krishna and Weller, Adrian},
	booktitle = {International Conference on Machine Learning},
	organization = {PMLR},
	pages = {16368--16382},
	title = {Measuring representational robustness of neural networks through shared invariances},
	year = {2022}}

@inproceedings{nguyen2021do,
	author = {Thao Nguyen and Maithra Raghu and Simon Kornblith},
	booktitle = {International Conference on Learning Representations},
	title = {Do Wide and Deep Networks Learn the Same Things? Uncovering How Neural Network Representations Vary with Width and Depth},
	url = {https://openreview.net/forum?id=KJNcAkY8tY4},
	year = {2021}}

@inproceedings{morcos2018insights,
	address = {Red Hook, NY, USA},
	author = {Morcos, Ari S. and Raghu, Maithra and Bengio, Samy},
	booktitle = {Proceedings of the 32nd International Conference on Neural Information Processing Systems},
	location = {Montr\'{e}al, Canada},
	numpages = {10},
	pages = {5732--5741},
	publisher = {Curran Associates Inc.},
	series = {NIPS'18},
	title = {Insights on representational similarity in neural networks with canonical correlation},
	year = {2018}}

@article{klabunde2025similarity,
	address = {New York, NY, USA},
	author = {Klabunde, Max and Schumacher, Tobias and Strohmaier, Markus and Lemmerich, Florian},
	doi = {10.1145/3728458},
	journal = {ACM Computing Surveys},
	number = {9},
	publisher = {Association for Computing Machinery},
	title = {Similarity of Neural Network Models: A Survey of Functional and Representational Measures},
	volume = {57},
	year = {2025}}

@article{ding2021grounding,
	author = {Ding, Frances and Denain, Jean-Stanislas and Steinhardt, Jacob},
	journal = {Advances in Neural Information Processing Systems},
	pages = {1556--1568},
	title = {Grounding representation similarity through statistical testing},
	volume = {34},
	year = {2021}}

@article{hryniowski2020inter,
	author = {Hryniowski, Andrew and Wong, Alexander},
	journal = {arXiv preprint arXiv:2012.03793},
	title = {Inter-layer information similarity assessment of deep neural networks via topological similarity and persistence analysis of data neighbour dynamics},
	year = {2020}}

@article{wang2020towards,
	author = {Wang, Chenxu and Rao, Wei and Guo, Wenna and Wang, Pinghui and Liu, Jun and Guan, Xiaohong},
	doi = {10.1109/TKDE.2020.2989512},
	journal = {IEEE Transactions on Knowledge and Data Engineering},
	number = {2},
	pages = {927--941},
	publisher = {IEEE},
	title = {Towards understanding the instability of network embedding},
	volume = {34},
	year = {2020}}

@inproceedings{schumacher2021effects,
	author = {Schumacher, Tobias and Wolf, Hinrikus and Ritzert, Martin and Lemmerich, Florian and Grohe, Martin and Strohmaier, Markus},
	booktitle = {Joint European Conference on Machine Learning and Knowledge Discovery in Databases},
	doi = {10.48550/arXiv.2005.10039},
	organization = {Springer},
	pages = {197--215},
	title = {The effects of randomness on the stability of node embeddings},
	year = {2021}}

@article{williams2021generalized,
	author = {Williams, Alex H and Kunz, Erin and Kornblith, Simon and Linderman, Scott},
	journal = {Advances in neural information processing systems},
	pages = {4738--4750},
	title = {Generalized shape metrics on neural representations},
	volume = {34},
	year = {2021}}

@article{shahbazi2021using,
	author = {Shahbazi, Mahdiyar and Shirali, Ali and Aghajan, Hamid and Nili, Hamed},
	doi = {10.1101/2020.11.25.398511},
	journal = {NeuroImage},
	pages = {118271},
	publisher = {Elsevier},
	title = {Using distance on the Riemannian manifold to compare representations in brain and in models},
	volume = {239},
	year = {2021}}

@inproceedings{kornblith2019similarity,
	author = {Kornblith, Simon and Norouzi, Mohammad and Lee, Honglak and Hinton, Geoffrey},
	booktitle = {International conference on machine learning},
	organization = {PMlR},
	pages = {3519--3529},
	title = {Similarity of neural network representations revisited},
	year = {2019}}

@inproceedings{gwilliam2022beyond,
	author = {Gwilliam, Matthew and Shrivastava, Abhinav},
	booktitle = {Proceedings of the IEEE/CVF conference on computer vision and pattern recognition},
	pages = {9642--9652},
	title = {Beyond supervised vs. unsupervised: Representative benchmarking and analysis of image representation learning},
	year = {2022}}

@article{yu2026parallel,
	author = {Yu, Zehua and Li, Xiaoyi and Liu, Pengfei and Tao, Jun},
	doi = {10.1109/TVCG.2026.3654590},
	journal = {IEEE Transactions on Visualization and Computer Graphics},
	number = {3},
	pages = {2758-2772},
	title = {{Parallel Clusters}: Visual Comparison of Embeddings Based on Multi-Scale Neighborhood Analysis},
	volume = {32},
	year = {2026}}

@inproceedings{li2018embeddingvis,
	author = {Li, Quan and Njotoprawiro, Kristanto Sean and Haleem, Hammad and Chen, Qiaoan and Yi, Chris and Ma, Xiaojuan},
	booktitle = {2018 IEEE Conference on Visual Analytics Science and Technology (VAST)},
	doi = {10.1109/VAST.2018.8802454},
	organization = {IEEE},
	pages = {48--59},
	title = {{EmbeddingVis}: A visual analytics approach to comparative network embedding inspection},
	year = {2018}}

@article{sucholutsky2025getting,
	author = {Sucholutsky, Ilia and Muttenthaler, Lukas and Weller, Adrian and Peng, Andi and Bobu, Andreea and Kim, Been and Love, Bradley C and Cueva, Christopher J and Grant, Erin and Groen, Iris and others},
	journal = {Transactions on Machine Learning Research},
	publisher = {Transactions on Machine Learning Research},
	title = {Getting aligned on representational alignment},
	volume = {2025},
	year = {2025}}

@article{rissom2024decoding,
	author = {Rissom, Pia Francesca and Sarmiento, Paulo Yanez and Safer, Jordan and Coley, Connor W and Renard, Bernhard Y and Heyne, Henrike O and Iqbal, Sumaiya},
	doi = {10.1101/2024.06.21.600139},
	journal = {bioRxiv},
	pages = {2024--06},
	publisher = {Cold Spring Harbor Laboratory},
	title = {Decoding protein language models: insights from embedding space analysis},
	year = {2024}}

@article{yan2025explainable,
	author = {Yan, Xinyuan and Sevastjanova, Rita and van der Ben, Sinie and El-Assady, Mennatallah and Wang, Bei},
	journal = {arXiv preprint arXiv:2507.18607},
	title = {{Explainable Mapper}: Charting LLM embedding spaces using perturbation-based explanation and verification agents},
	year = {2025}}

@inproceedings{radford2021learning,
	author = {Radford, Alec and Kim, Jong Wook and Hallacy, Chris and Ramesh, Aditya and Goh, Gabriel and Agarwal, Sandhini and Sastry, Girish and Askell, Amanda and Mishkin, Pamela and Clark, Jack and others},
	booktitle = {International conference on machine learning},
	organization = {PMLR},
	pages = {8748--8763},
	title = {Learning transferable visual models from natural language supervision},
	year = {2021}}

@inproceedings{lin2014microsoft,
	author = {Lin, Tsung-Yi and Maire, Michael and Belongie, Serge and Hays, James and Perona, Pietro and Ramanan, Deva and Doll{\'a}r, Piotr and Zitnick, C Lawrence},
	booktitle = {European conference on computer vision},
	doi = {10.1007/978-3-319-10602-1_48},
	organization = {Springer},
	pages = {740--755},
	title = {{Microsoft COCO}: Common objects in context},
	year = {2014}}

@inproceedings{devlin2019bert,
	author = {Jacob Devlin and Ming{-}Wei Chang and Kenton Lee and Kristina Toutanova},
	booktitle = {Proceedings of the 2019 Conference of the North American Chapter of the Association for Computational Linguistics: Human Language Technologies, Volume 1 (Long and Short Papers)},
	doi = {10.18653/V1/N19-1423},
	pages = {4171--4186},
	publisher = {Association for Computational Linguistics},
	title = {{BERT:} Pre-training of Deep Bidirectional Transformers for Language Understanding},
	url = {https://doi.org/10.18653/v1/n19-1423},
	year = {2019}}

@inproceedings{jiao2020tinybert,
	address = {Online},
	author = {Jiao, Xiaoqi and Yin, Yichun and Shang, Lifeng and Jiang, Xin and Chen, Xiao and Li, Linlin and Wang, Fang and Liu, Qun},
	booktitle = {Findings of the Association for Computational Linguistics: EMNLP 2020},
	doi = {10.18653/v1/2020.findings-emnlp.372},
	editor = {Cohn, Trevor and He, Yulan and Liu, Yang},
	month = nov,
	pages = {4163--4174},
	publisher = {Association for Computational Linguistics},
	title = {{T}iny{BERT}: Distilling {BERT} for Natural Language Understanding},
	url = {https://aclanthology.org/2020.findings-emnlp.372/},
	year = {2020}}

@inproceedings{schneider2015corpus,
	author = {Schneider, Nathan and Smith, Noah A},
	booktitle = {Proceedings of the 2015 Conference of the North American Chapter of the Association for Computational Linguistics: Human Language Technologies},
	doi = {10.3115/v1/N15-1177},
	pages = {1537--1547},
	title = {A corpus and model integrating multiword expressions and supersenses},
	year = {2015}}

@article{SaggarSpornsGonzalez-Castillo2018,
	author = {Manish Saggar and Olaf Sporns and Javier Gonzalez-Castillo and Peter A. Bandettini and Gunnar Carlsson and Gary Glover and Allan L. Reiss},
	doi = {10.1038/s41467-018-03664-4},
	journal = {Nature Communications},
	number = {1399},
	title = {Towards a new approach to reveal dynamical organization of the brain using topological data analysis},
	volume = {9},
	year = {2018}}

@article{GeniesseSpornsPetri2019,
	author = {Caleb Geniesse and Olaf Sporns and Giovanni Petri and Manish Saggar},
	doi = {10.1162/netn_a_00093},
	journal = {Network Neuroscience},
	number = {3},
	title = {Generating dynamical neuroimaging spatiotemporal representations {(DyNeuSR)} using topological data analysis},
	volume = {3},
	year = {2019}}

@article{Knudson2020,
	author = {Ang\'{e}lica Knudson and Felipe Gonz\'{a}lez-Casabianca and Alejandro Feged-Rivadeneira and Maria Fernanda Pedreros and Samanda Aponte and Adriana Olaya and Carlos F. Castillo and Elvira Mancilla and Anderson Piamba-Dorado and Ricardo Sanchez-Pedraza and Myriam Janeth Salazar-Terreros and Naomi Lucchi and Venkatachalam Udhayakumar and Chris Jacob and Alena Pance and Manuela Carrasquilla and Giovanni Apr\'{a}ez and Jairo Andr\'{e}s Angel and Julian C. Rayner and Vladimir Corredor},
	doi = {10.1038/s41598-020-60676-1},
	journal = {Scientific Reports},
	number = {3756},
	title = {Spatio-temporal dynamics of Plasmodium falciparum transmission within a spatial unit on the Colombian Pacific Coast},
	volume = {10},
	year = {2020}}

@article{JeitzinerCarriereRougemont2019,
	author = {Rachel Jeitziner and Mathieu Carri\'{e}re and Jacques Rougemont and Steve Oudot and Kathryn Hess and Cathrin Brisken},
	doi = {10.1093/bioinformatics/btz052},
	journal = {Bioinformatics},
	number = {18},
	pages = {3339-3347},
	title = {Two-Tier Mapper, an unbiased topology-based clustering method for enhanced global gene expression analysis},
	volume = {35},
	year = {2019}}

@article{Alagappan2012,
	author = {Muthu Alagappan},
	journal = {MIT Sloan Sports Analytics Conference},
	title = {From 5 to 13: Redefining the Positions in Basketball},
	year = {2012}}

@article{MathewsNadeemLevine2019,
	author = {James C. Mathews and Saad Nadeem and Arnold J. Levine and Maryam Pouryahya and Joseph O. Deasy and Allen Tannenbaum},
	doi = {10.1038/s41523-019-0124-8},
	journal = {NPJ Breast Cancer},
	number = {30},
	title = {Robust and interpretable {PAM50} reclassification exhibits survival advantage for myoepithelial and immune phenotypes},
	volume = {5},
	year = {2019}}

@article{PataniaVaccarinoPetri2017,
	author = {Alice Patania and Francesco Vaccarino and Giovanni Petri},
	doi = {10.1140/epjds/s13688-017-0104-x},
	journal = {EPJ Data Science},
	number = {7},
	title = {Topological analysis of data},
	volume = {6},
	year = {2017}}

@article{ZhouZhouDing2023,
	author = {Youjia Zhou and Yi Zhou and Jie Ding and Bei Wang},
	journal = {Topology, Algebra, and Geometry in Machine Learning (TAGML) Workshop at ICML},
	title = {Visualizing and Analyzing the Topology of Neuron Activations in Deep Adversarial Training},
	year = {2023}}

@article{NicolauLevineCarlsson2011,
	author = {Nicolau, Monica and Levine, Arnold J and Carlsson, Gunnar},
	doi = {10.1073/pnas.1102826108},
	journal = {Proceedings of the National Academy of Sciences},
	number = {17},
	pages = {7265--7270},
	title = {Topology based data analysis identifies a subgroup of breast cancers with a unique mutational profile and excellent survival},
	volume = {108},
	year = {2011}}

@article{rathore2023topobert,
	author = {Rathore, Archit and Zhou, Yichu and Srikumar, Vivek and Wang, Bei},
	doi = {10.1177/14738716231168671},
	journal = {Information Visualization},
	number = {3},
	pages = {186-208},
	publisher = {SAGE Publications Sage UK: London, England},
	title = {{TopoBERT}: Exploring the topology of fine-tuned word representations},
	volume = {22},
	year = {2023}}

@inproceedings{ester1996density,
	author = {Ester, Martin and Sander, Joerg and Xu, Xiaowei},
	booktitle = {Proceedings of the 2nd International Conference on Knowledge Discovery and Data Mining},
	pages = {226--231},
	title = {A density-based algorithm for discovering clusters in large spatial databases with noise},
	volume = {96},
	year = {1996}}

@inproceedings{xenopoulos2022gale,
	author = {Xenopoulos, Peter and Chan, Gromit and Doraiswamy, Harish and Nonato, Luis Gustavo and Barr, Brian and Silva, Claudio},
	booktitle = {Topological, Algebraic and Geometric Learning Workshops 2022},
	organization = {PMLR},
	pages = {322--331},
	title = {{GALE}: Globally assessing local explanations},
	year = {2022}}

@inproceedings{rair2025annotators,
	author = {Rair, Nisrine and Goupil, Alban and Vrabie, Valeriu and Chochoy, Emmanuel},
	booktitle = {Proceedings of the 2025 Conference on Empirical Methods in Natural Language Processing},
	doi = {10.18653/v1/2025.emnlp-main.426},
	pages = {8468--8491},
	title = {When Annotators Disagree, Topology Explains: Mapper, a Topological Tool for Exploring Text Embedding Geometry and Ambiguity},
	year = {2025}}

@article{heimerl2020embcomp,
	author = {Heimerl, Florian and Kralj, Christoph and Moller, Torsten and Gleicher, Michael},
	doi = {10.1109/TVCG.2020.3045918},
	journal = {{IEEE} Transactions on Visualization and Computer Graphics},
	number = {8},
	pages = {2953-2969},
	title = {{embComp}: Visual Interactive Comparison of Vector Embeddings},
	volume = {28},
	year = {2020}}

@article{xuan2022vac,
	author = {Xuan, Xiwei and Zhang, Xiaoyu and Kwon, Oh-Hyun and Ma, Kwan-Liu},
	doi = {10.1109/TVCG.2022.3165347},
	journal = {IEEE Transactions on Visualization and Computer Graphics},
	number = {6},
	pages = {2326--2337},
	publisher = {IEEE},
	title = {{VAC-CNN}: A visual analytics system for comparative studies of deep convolutional neural networks},
	volume = {28},
	year = {2022}}

@article{solunke2024mountaineer,
	author = {Solunke, Parikshit and Guardieiro, Vitoria and Rulff, Joao and Xenopoulos, Peter and Chan, Gromit Yeuk-Yin and Barr, Brian and Nonato, Luis Gustavo and Silva, Claudio},
	doi = {10.1109/TVCG.2024.3418653},
	journal = {IEEE Transactions on Visualization and Computer Graphics},
	number = {12},
	pages = {7763--7775},
	publisher = {IEEE},
	title = {Mountaineer: Topology-driven visual analytics for comparing local explanations},
	volume = {30},
	year = {2024}}

@incollection{shneiderman2003eyes,
	author = {Shneiderman, Ben},
	booktitle = {The craft of information visualization},
	doi = {10.1109/VL.1996.545307},
	pages = {364--371},
	publisher = {Elsevier},
	title = {The eyes have it: A task by data type taxonomy for information visualizations},
	year = {2003}}

@article{rathore2021topoact,
	author = {Rathore, Archit and Chalapathi, Nithin and Palande, Sourabh and Wang, Bei},
	doi = {10.1111/cgf.14195},
	journal = {Computer Graphics Forum},
	number = {1},
	pages = {382-397},
	title = {{TopoAct}: Visually Exploring the Shape of Activations in Deep Learning},
	volume = {40},
	year = {2021}}

@article{SinghMemoliCarlsson2007,
	author = {Singh, Gurjeet and M{\'e}moli, Facundo and Carlsson, Gunnar E},
	doi = {10.2312/SPBG/SPBG07/091-100},
	journal = {Eurographics Symposium on Point-Based Graphics},
	pages = {91-100},
	title = {Topological methods for the analysis of high dimensional data sets and {3D} object recognition},
	year = {2007}}

@article{sevastjanova2022lmfingerprints,
	author = {Sevastjanova, Rta and Kalouli, Aikaterini-Lida and Beck, Christin and Hauptmann, Hanna and El-Assady, Menna},
	doi = {https://doi.org/10.1111/cgf.14541},
	eprint = {https://onlinelibrary.wiley.com/doi/pdf/10.1111/cgf.14541},
	journal = {Computer Graphics Forum},
	number = {3},
	pages = {295-307},
	title = {LMFingerprints: Visual Explanations of Language Model Embedding Spaces through Layerwise Contextualization Scores},
	url = {https://onlinelibrary.wiley.com/doi/abs/10.1111/cgf.14541},
	volume = {41},
	year = {2022}}

@article{ye2026dkmap,
	author = {Ye, Yilin and Ruan, Chenxi and Zhang, Yu and Deng, Zikun and Zeng, Wei},
	doi = {10.1109/TVCG.2025.3642641},
	journal = {IEEE Transactions on Visualization and Computer Graphics},
	month = {01},
	pages = {1-11},
	title = {{DKMap}: Interactive Exploration of Vision-Language Alignment in Multimodal Embeddings via Dynamic Kernel Enhanced Projection},
	volume = {PP},
	year = {2026}}

@article{lee2026unlearning,
	author = {Lee, Jaeung and Yu, Suhyeon and Jang, Yurim and Woo, Simon S and Jo, Jaemin},
	doi = {10.1109/TVCG.2026.3658325},
	journal = {IEEE Transactions on Visualization and Computer Graphics},
	number = {3},
	pages = {2852-2867},
	publisher = {IEEE},
	title = {{Unlearning Comparator}: A Visual Analytics System for Comparative Evaluation of Machine Unlearning Methods},
	volume = {32},
	year = {2026}}

@article{sevastjanova2022adapters,
	author = {Sevastjanova, Rita and Cakmak, Eren and Ravfogel, Shauli and Cotterell, Ryan and El-Assady, Mennatallah},
	doi = {10.1109/TVCG.2022.3209458},
	journal = {IEEE Transactions on Visualization and Computer Graphics},
	number = {1},
	pages = {1178--1188},
	publisher = {IEEE},
	title = {Visual Comparison of Language Model Adaptation},
	volume = {29},
	year = {2022}}

@inproceedings{romero2024resilient,
	author = {Romero-Alvarado, Daniel and Hern{\'a}ndez-Orallo, Jos{\'e} and Mart{\'\i}nez-Plumed, Fernando},
	booktitle = {International Conference on Intelligent Data Engineering and Automated Learning},
	doi = {10.1007/978-3-031-77731-8_8},
	pages = {85--96},
	title = {How Resilient are Language Models to Text Perturbations?},
	year = {2024}}

@inproceedings{sevastjanova2021explaining,
	author = {Rita Sevastjanova and Aikaterini-Lida Kalouli and Christin Beck and Hannah Hauptmann and Mennatallah El-Assady},
	booktitle = {Proc. of the Association for Computational Linguistics},
	doi = {10.48448/1bf4-bg31},
	location = {Bangkok, Thailand},
	publisher = {ACL},
	series = {ACL},
	title = {{Explaining Contextualization in Language Models using Visual Analytics}},
	year = {2021}}

@inproceedings{sivaraman2022emblaze,
	author = {Sivaraman, Venkatesh and Wu, Yiwei and Perer, Adam},
	booktitle = {27th Int. Conf. on Intelligent User Interfaces},
	doi = {https://doi.org/10.1145/3490099.3511137},
	pages = {418--432},
	title = {Emblaze: Illuminating machine learning representations through interactive comparison of embedding spaces},
	year = {2022}}
\newpage
\appendix 
\crefalias{section}{appendix} 

In this supplement, we first provide details on the spectral clustering procedure and the \textit{structural coherence} metric used to evaluate each local alignment pair (\cref{supp:local-align-details}). We then describe the fine-tuning procedures for the language models used in~\cref{sec:LLM-case} (\cref{supp:fine-tuning-details}). Finally, we present additional mapper graphs comparing layers 7 and 12 of the fine-tuned BERT-base model, as discussed in~\cref{sec:layer-compare} (\cref{supp:layer7vs12}).

\section{Implementation Details for Local Alignment}
\label{supp:local-align-details}

\para{Spectral clustering.}  
As described in~\cref{sec:alignment_detection}, we apply spectral clustering to the joint mapper graph to identify local alignments between two mapper graphs $G_1$ and $G_2$:
\[
G_{\text{joint}} = (V, E) = (V_1 \cup V_2,\; E_1 \cup E_2 \cup E_{\text{inter}}).
\]
Let $W \in \mathbb{R}^{|V| \times |V|}$ denote the affinity matrix, defined as
\[
W_{ij} =
\begin{cases}
\alpha, & \text{if } (i,j) \in E_1, \\
\beta, & \text{if } (i,j) \in E_2, \\
\gamma, & \text{if } (i,j) \in E_{\text{inter}},
\end{cases}
\]
where $\alpha$, $\beta$, and $\gamma$ control the relative importance of intra-graph and inter-graph edges. By default, we set $\alpha = \beta = \gamma = 1$, treating all edges uniformly so that clustering is driven primarily by graph topology.

We then construct the normalized graph Laplacian:
\[
L = I - D^{-1/2} W D^{-1/2},
\]
where $D$ is the diagonal degree matrix with entries
\[
D_{ii} = \sum_j W_{ij}.
\]
Next, we compute the first $k$ eigenvectors of $L$ to obtain a $k$-dimensional spectral embedding. Each node $v \in V$ is represented by a vector $\mathbf{z}_v \in \mathbb{R}^k$, corresponding to the $v$-th row of the eigenvector matrix. We then apply $k$-means clustering to these embeddings to partition the nodes into local alignment groups.

The number of clusters $k$ is selected heuristically from the eigenspectrum by identifying the first elbow point in the sorted eigenvalues.

\para{Structural coherence computation.}  
As described in~\cref{sec:alignment_detection}, we quantify the structural coherence of each local alignment pair (i.e., each cluster obtained through spectral clustering) using the Silhouette score:
\[
\text{Silhouette}(v) = \frac{b(v) - a(v)}{\max\{a(v), b(v)\}},
\]
where $a(v)$ denotes the average distance between node $v$ and all other nodes within the same cluster, and $b(v)$ denotes the average distance between $v$ and nodes in the nearest neighboring cluster.

Distances are computed in the spectral embedding space using the corresponding node representations $\mathbf{z}_v$. We then aggregate the Silhouette scores across all nodes within a cluster to obtain a single structural coherence score for each local alignment pair.

\begin{table}[h]
\centering
\begin{tabular}{lccc}
\toprule
\textbf{Model} & \textbf{Layers} & \textbf{Hidden Dim} & \textbf{Params} \\
\midrule
BERT-base & 12 & 768 & 110M \\
BERT-tiny & 2 & 128 & 4.4M \\
\bottomrule
\end{tabular}
\caption{Summary of the language models used in the fine-tuning tasks. The corresponding HuggingFace model identifiers are \texttt{google-bert/bert-base-uncased} for BERT-base and \texttt{prajjwal1/bert-tiny} for BERT-Tiny.}
\label{tab:model-summary}
\end{table}

\section{Fine-Tuning Details}
\label{supp:fine-tuning-details}

In~\cref{sec:LLM-case}, we fine-tune BERT-base and BERT-Tiny on the STREUSLE v4.2 dataset for a word classification task based on Supersense Role labels. Both models are fine-tuned using the Hugging Face library with the AdamW optimizer and a batch size of 32. We employ a linear learning-rate scheduler with a 10\% warm-up phase and a fixed learning rate of $3 \times 10^{-4}$. Each model is trained for six epochs, with checkpoints saved after every epoch.

After fine-tuning, we extract contextualized embeddings for 4,872 annotated words from the training set at every epoch across all model layers. \cref{tab:model-summary} summarizes the number of layers, embedding dimensions, model sizes, and Hugging Face model identifiers for each pretrained model.

\section{Comparison of Layers 7 and 12 in the Fine-Tuned BERT-Base Model}
\label{supp:layer7vs12}

\begin{figure}[!t]
    \centering
    \includegraphics[width=\linewidth]{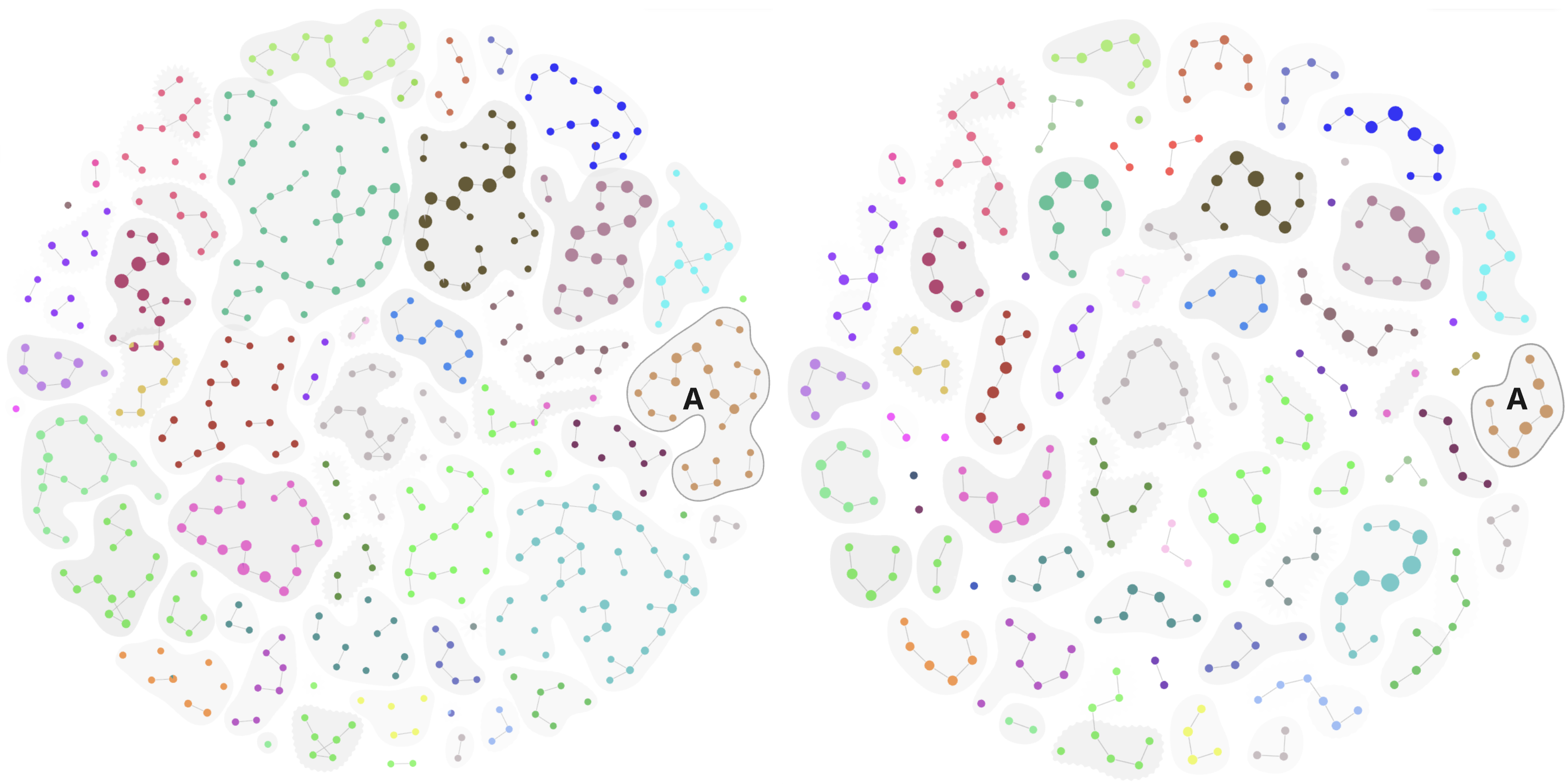}
    \vspace{-2em}
    \caption{Mapper graphs of layers 7 (left) and 12 (right) of the fine-tuned BERT-base model.}
    \label{fig:layer7-12}
    \vspace{-2em}
\end{figure}

\cref{fig:layer7-12} presents the mapper graphs for layers 7 (left) and 12 (right) of the fine-tuned BERT-base model, with local alignments visualized using Bubble Sets. The highlighted local alignment pair A corresponds to the example discussed in Section~\cref{sec:layer-compare}.

\end{document}